 \pdfoutput=1
 \documentclass[preprint,A4]{elsarticle}

\usepackage[width=15cm,height=22cm,left=2.75cm, top=2.75cm]{geometry}
\usepackage{placeins}  
\usepackage{float}     
\usepackage{verbatim}  
\usepackage{xspace}    
\usepackage{changepage} 

\usepackage{subcaption}
\usepackage{booktabs} 

\usepackage[bookmarks=false]{hyperref}
\usepackage{glossaries} 

\usepackage{amsmath} 
\usepackage{amssymb} 

\journal{Journal of \LaTeX\ Templates}

\newacronym{CR}{CR}{Cosmic Ray}
\newacronym{CRs}{CRs}{Cosmic Rays}
\newacronym{UHECR}{UHECR}{Ultra High Energy Cosmic Rays}
\newacronym{UHE}{UHE}{Ultra High Energy}
\newacronym{SEP}{SEP}{solar energetic particles}
\newacronym{GCR}{GCR}{Galactic Cosmic Rays}
\newacronym{ACR}{ACR}{Anomalous cosmic rays}
\newacronym{GZK}{GZK}{Greisen-Zatsepin-Kuzmin}
\newacronym{CMB}{CMB}{Cosmic Microwave Background Radiation}
\newacronym{UV}{UV}{Ultra-Violet radiation}
\newacronym{IR}{IR}{Infra-Red radiation}
\newacronym{Opt}{Opt}{Optical radiation}
\newacronym{URB}{URB}{Cosmic Universal Radio Radiation} 
\newacronym{SNR}{SNRs}{Supernova Remnants}
\newacronym{AGN}{AGN}{Active Galactic Nuclei}
\newacronym{VCV}{VCV}{V\'{e}ron-Cetty and V\'{e}ron catalogue}
\newacronym{ICS}{ICS}{Inverse Compton Scattering}

\newacronym{RFT}{RFT}{Gribov's Reggeon Field Theory}
\newacronym{LPM}{LPM}{Landau-Pomeranchuk-Migdal effect}
\newacronym{QCD}{QCD}{Quantum Chromodynamics}
\newacronym{DPM}{DPM}{Dual Parton Model}
\newacronym{EPOS}{E{\textsc{pos}}}{Energy conserving quantum mechanical multiple scattering approach, based on Partons, Off-shell remnants and Splitting of partons ladders}
\newacronym{QGSjet}{QGSJ{\textsc{et}}}{Quark Gluon String Model with mini-Jet}
\newacronym{Fluka}{FLUKA}{Gamma-Hadron-Electron-Interaction SH(A)ower code}
\newacronym{Gheisha}{Gheisha}{FLUktuierende KAskade (German for 
fluctuating cascade)}
\newacronym{CORSIKA}{CORSIKA}{COsmic Ray SImulations for KAscade}
\newacronym{AIRES}{AIRES}{AIRshower Extended Simulations}

\newacronym{CERN}{CERN}{European Organization for Nuclear Research (a particle physics laboratory in Geneva, Switzerland)}
\newacronym{LHC}{LHC}{Large Hadron Collider}
\newacronym{RHIC}{RHIC}{Relativistic Heavy Ion Collider (at BNL)}
\newacronym{TOTEM}{TOTEM}{TOTal Elastic and diffractive cross section Measurement}

\newacronym{KASCADE}{KASCADE}{KArlsruhe Shower Core and Array DEtector}
\newacronym{HiRes}{HiRes}{High Resolution Fly's Eye detector}
\newacronym{TA}{TA}{Telescope Array}
\newacronym{AGASA}{AGASA}{Akeno Giant Air Shower Array}

\newacronym{PAO}{PAO}{Pierre Auger Observatory}
\newacronym{HEAT}{HEAT}{High Elevation Auger Telescope}
\newacronym{AMIGA}{AMIGA}{Auger Muons and Infill for the Ground Array}
\newacronym{AERA}{AERA}{Auger Engineering Radio Array}
\newacronym{MIDAS}{MIDAS}{MIcrowave Detection of Air Showers}
\newacronym{AMBER}{AMBER}{Air-shower Microwave Bremsstrahlung Experimental Radiometer}
\newacronym{EASIER}{EASIER}{Extensive Air Shower Identification using Electron Radiometers}
\newacronym{LOPES}{LOPES}{LOFAR Prototype Station}
\newacronym{LOFAR}{LOFAR}{Low Frequency ARray}
\newacronym{CROME}{CROME}{Cosmic Ray Observation via Microwave Emission}
\newacronym{CODALEMA}{CODALEMA}{COsmic ray Detection Array with Logarithmic ElectroMagnetic Antennas}
\newacronym{AIRFLY}{AIRFLY}{AIR FLuorescence Yield}

\newacronym{rpc}{RPC}{Resistive Plate Chambers}
\newacronym{marta}{MARTA}{Muon Auger RPC for the Tank Array}

\newacronym{CTA}{CTA}{Cherenkov Telescope Array}
\newacronym{MAGIC}{MAGIC}{Major Atmospheric Gamma-ray Imaging Cherenkov}
\newacronym{HESS}{HESS}{High Energy Stereoscopic System}
\newacronym{VERITAS}{VERITAS}{Very Energetic Radiation Imaging Telescope Array System}
\newacronym{HAWC}{HAWC}{the High-Altitude Water Cherenkov Observatory}
\newacronym{SUGAR}{SUGAR}{The Sydney University array}
\newacronym{MACFLY}{MACFLY}{Measurement of Air Cherenkov and Fluorescence Light Yield}

\newacronym{EAS}{EAS}{Extensive Air Shower}
\newacronym{USP}{USP}{Universal Shower Profile}
\newacronym{ldf}{LDF}{Lateral Distribution Function}
\newacronym{MC}{MC}{Monte Carlo}
\newacronym{MPD}{MPD}{Muon Production Depth}
\newacronym{VEM}{VEM}{Vertical Equivalent Muon}
\newacronym{em}{EM}{Electromagnetic signal component}
\newacronym{mu}{MU}{Muonic signal component}
\newacronym{tot}{TOT}{Total signal component}
\newacronym{CIC}{CIC}{Constant Intensity Cut}
\newacronym{NKG}{NKG}{Nishimura, Kamata and Greisen equation}
\newacronym{FLY}{FLY}{Fluorescence Light Yield}

\newacronym{wct}{WCT}{Water-Cherenkov tanks} 
\newacronym{wcd}{WCD}{Water Cherenkov detector} 
\newacronym{PMT}{PMT}{Photomultiplier Tube}
\newacronym{sd}{SD}{Surface Detectors}
\newacronym{fd}{FD}{Fluorescence Detectors}
\newacronym{fov}{FOV}{Field Of View}
\newacronym{SDP}{SDP}{Shower Detector Plane}
\newacronym{FADC}{FADC}{Flash Analog to Digital Converter}
\newacronym{ADC}{ADC}{Analog Digital Converter}
\newacronym{GPS}{GPS}{Global Positioning System}
\newacronym{DAQ}{DAQ}{Data Acquisition}
\newacronym{XML}{XML}{eXtensible Markup Language}

\newacronym{CLF}{CLF}{Central Laser Facility}
\newacronym{XLF}{XLF}{Extreme Laser Facility}
\newacronym{cdas}{CDAS}{Central Data Acquisition System}
\newacronym{LIDAR}{LIDAR}{Light Detection And Ranging}
\newacronym{ham}{HAM}{Horizontal Attenuation Monitors}
\newacronym{FRAM}{FRAM}{tttttt}
\newacronym{fram0}{FRAM}{The Photometric Robotic Atmospheric Monitor}
\newacronym{apf}{APF}{Aerosol Phase Function}
\newacronym{FPGA}{FPGA}{Field-Programmable Gate Array}

\newacronym{UrQMD}{UrQMD}{Ultra-relativistic Quantum Molecular Dynamics model}
\newacronym{QGS}{QGS}{Quark-Gluon-String}
\newacronym{CL}{CL}{Confidence Level}

\newacronym{VHF}{VHF}{Very High Frequency radiation}
\newacronym{MBR}{MBR}{molecular bremsstrahlung radiation}
\newacronym{SOPHIA}{SOPHIA}{Simulations Of Photo Hadronic Interactions in Astrophysics}

\newacronym{PLD}{PLD}{Programmable Logic Device}
\newacronym{FE}{FE}{front-end}
\newacronym{CCD}{CCD}{charge-coupled device} 

\newacronym{FLT}{FLT}{First Level Trigger in FD}
\newacronym{SLT}{SLT}{Second Level Trigger in FD}
\newacronym{TLT}{TLT}{Third Level Trigger in FD}
\newacronym{T1}{T1}{First Level Trigger in SD} 
\newacronym{T2}{T2}{Second Level Trigger in SD} 
\newacronym{T3}{T3}{Third Level Trigger in SD} 
\newacronym{T4}{T4}{Fourth Level Trigger (of physics trigger) in SD} 
\newacronym{T5}{T5}{Fifth Level Trigger (or fiducial trigger) in SD} 
\newacronym{TH}{TH}{T1 Simple Threshold trigger in SD} 
\newacronym{ToT}{ToT}{Time-over-Threshold trigger in SD}

\newcommand{\qgs}{QGSJ{\textsc{et}}-II.04\xspace}

\newcommand{\GHEISHA}{GHEISHA}

\begin{document}

\begin{frontmatter}
\title{3D simulation for Cherenkov emission in Extensive Air Showers}

\author[LIPadress]{J. Espadanal \corref{mycorrespondingauthor}}
\cortext[mycorrespondingauthor]{Corresponding author}
\ead{jespada@lip.pt}
\author[LIPadress]{P. Gon\c{c}alves}
\ead{patricia@lip.pt}

\address[LIPadress]{LIP, Av. Elias Garcia, 14-1, 1000-149 Lisboa, Portugal}

\begin{abstract}
The development of a 3-dimensional simulation for Cherenkov photon emissions in Extensive Air Showers (EAS) is reported in this paper. CORSIKA is the most widely used Monte-Carlo generator for the description of EAS, but it is not recommended to calculate Cherenkov light emissions for EAS at ultra high energies due to the enormous amount of data storage and running time required.   
The presented BinTheSky is a framework to simulate the Cherenkov light emissions using the spatial information produced by Monte-Carlo generators. The light is emitted in the shower, propagated and attenuated to the ground.  
The framework enables one to calculate the light spatial, timing and directional distributions at the ground or at a given altitude, whereas the usual approach of Ultra High Energy Cosmic Ray experiments relies on the simulation of the longitudinal shower development and on parametrizations of the transverse shower distributions.
\end{abstract}


\end{frontmatter}

\section{Introduction}

The \gls{UHECR} reaching the Earth have extremely low fluxes and cannot be detected directly in space. These particles enter the atmosphere and interact with atmospheric molecules producing cascades of secondary particles known as \glspl{EAS}. EAS can be detected through two main techniques: by observing the light emitted along the development of the shower or by detecting the shower particles surviving to the ground. Regarding the first technique, the light emitted is essentially fluorescence light (from atmospheric  nitrogen) and Cherenkov light from the relativistic particles in the shower. 
One of the most used cascade generator is \gls{CORSIKA}\cite{CORSIKA}, which for lower primary particle energies ($ \lesssim 10^{15}$ eV), can also be used to calculate direct Cherenkov light emission. However, at ultra high energies, due to the increase in shower particle multiplicity, calculations of direct Cherenkov light emission are prohibitive in terms of computing time and data storage capacity.

Accurate simulations or parametrizations of transverse profiles of EAS are fundamental for cosmic ray experiments which detect the light emitted by EAS. The Pierre Auger Experiment\cite{PAO_0} and others, reconstruct EAS based on the longitudinal shower development leaving out information on transverse shower structure. This approach is valid for distant showers, however, in close-by events, it neglects the additional data that lateral shower profiles can provide. To recover the shower shape, lateral average parameterizations are usually considered (see for example \cite{Nerling2006,Gora}). These average distributions do not account for shower to shower fluctuations which can be important in the interpretation of the data. Moreover, accurate descriptions of transverse shower profiles is important in arrays of Cherenkov detectors, such as Tunka\cite{TunkaExp}, Yakutsk\cite{YakutskExp} and CTA\cite{CTA} experiments, to obtain the lateral profiles and also to obtain the time distribution of Cherenkov light at ground level which may be used to recover the longitudinal shower maximum.

The development of \textit{BinTheSky} framework was motivated to allow new studies such as the recovery of the longitudinal profile from arrival time distributions at ground and to provide a framework for the study of different aspects of Cherenkov and fluorescence light emission and their propagation in EAS. 

In section \ref{section:Framework}, the BinTheSky framework is described, along with the necessary CORSIKA intervention. The light emissions, propagation and attenuations are also discussed. In \ref{section:Corsika} the Cherenkov light distributions, from the BinTheSky framework, are compared to those obtained with CORSIKA for EAS with $10^{14}$ and $10^{15}$ eV primary energy. In section \ref{section:Ground}, the Cherenkov light distributions obtained with BinTheSky at ground level are shown, together with the parametrization of the lateral density of Cherenkov photons and with the corresponding timing distributions at ground level. The results obtained with BinTheSky and future prospects are presented in section \ref{section:Conclusions}.

\section{Framework and Method}
\label{section:Framework}
\begin{figure}[!hb]
\begin{center}\centering
       \begin{subfigure}[!hb]{0.30\textwidth}\centering
                \includegraphics[width=1\textwidth]{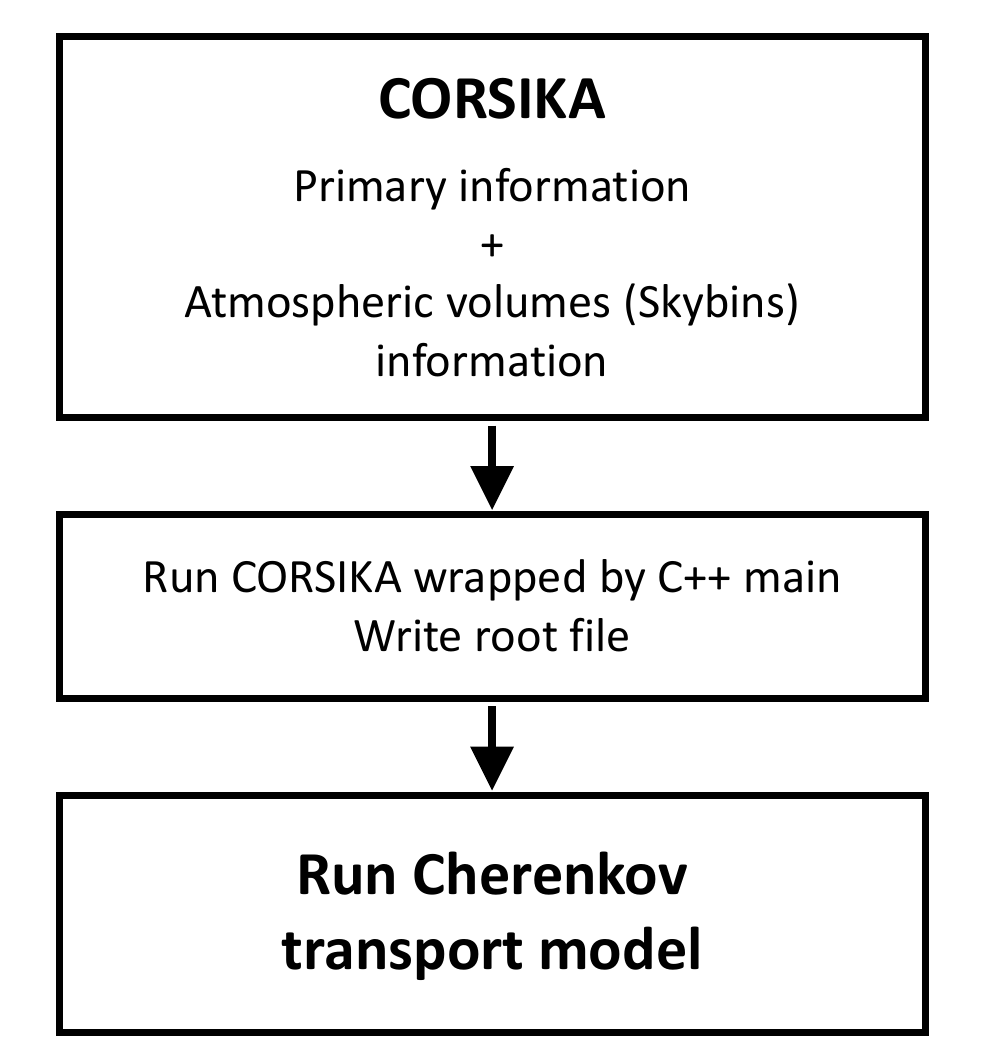}
                \caption{Framework data flow.}
                \label{fig: SkiBin a}
        \end{subfigure}%
       \begin{subfigure}[h]{0.35\textwidth}\centering
                \includegraphics[width=1.0\textwidth]{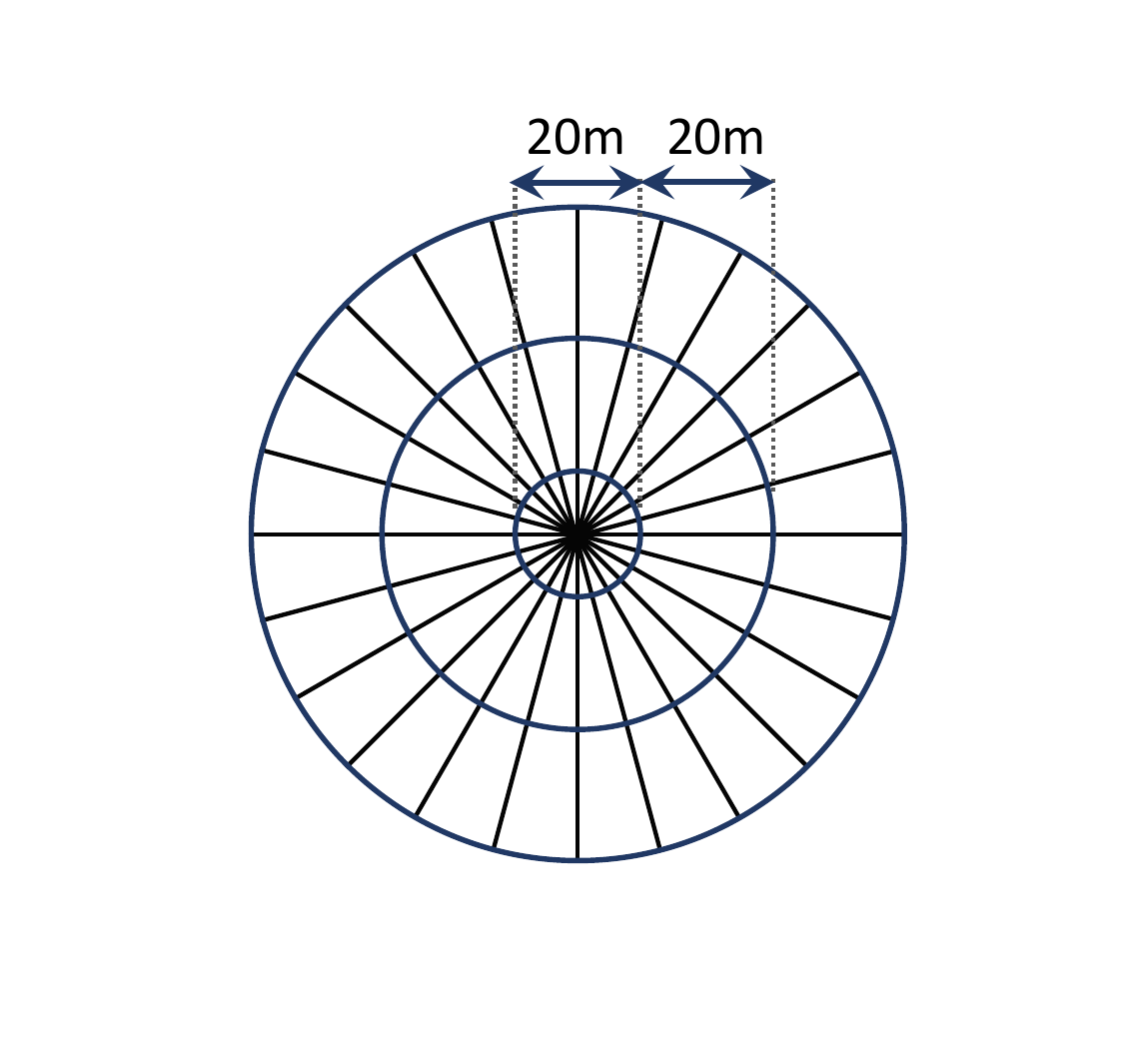}
                \caption{Transverse BinTheSky geometry}
                \label{fig: SkiBin b}
        \end{subfigure}%
       \begin{subfigure}[h]{0.30\textwidth}\centering
                \includegraphics[width=1.0\textwidth]{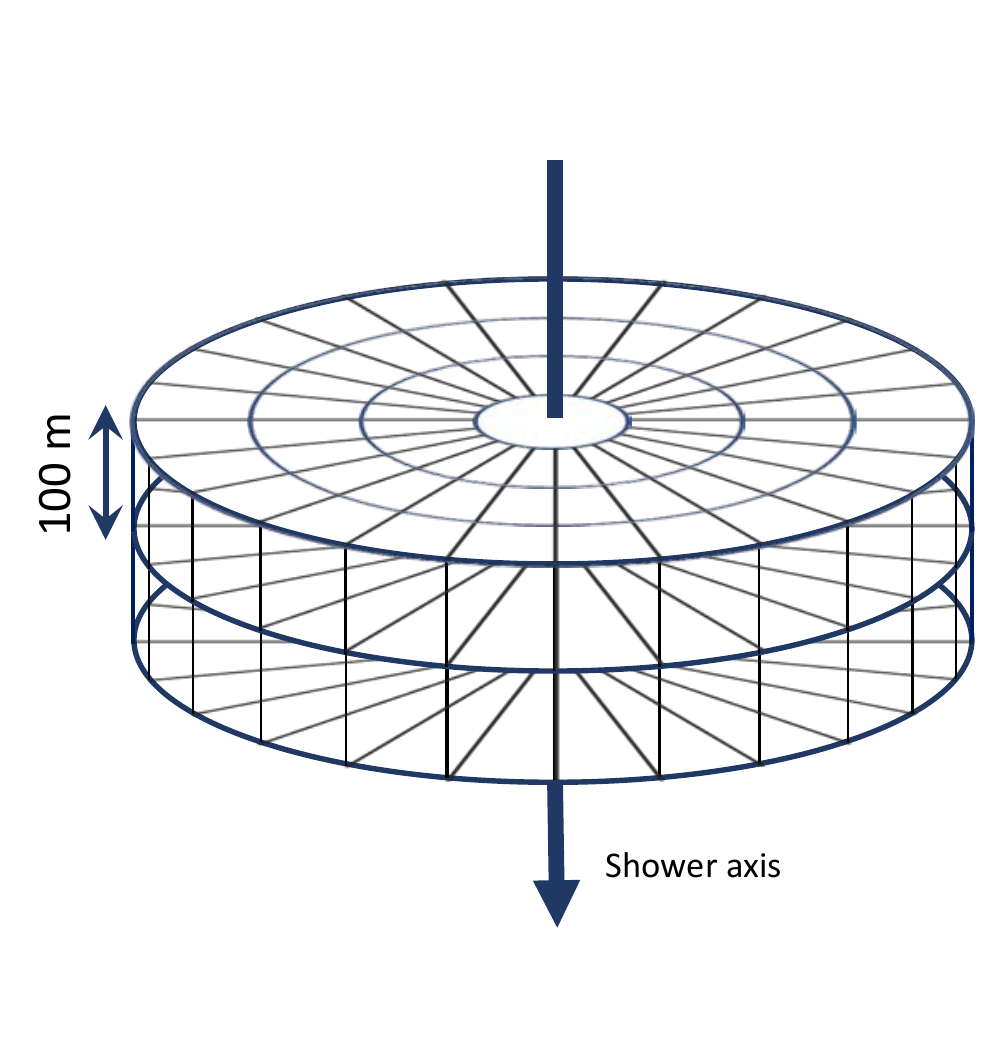}
                \caption{Two SkyBins layers}
                \label{fig: SkiBin c}
        \end{subfigure}%
\caption[The BinTheSky framework data flow and representation]{a) Representation of the data flow of the BinTheSky framework. Representation of the transverse structure of the BinTheSky geometry (b) and two SkyBins layers in the BinTheSky geometry (c), for a fixed z coordinate. The dark arrow represents the vertical shower axis, i.e. the direction of the primary particle.
}
\label{fig: SkiBin}
\end{center}
\end{figure}

The \textit{BinTheSky} framework has the purpose of saving the spatial information produced at generator level in EAS simulations, in a way, that it can be used to simulate the production of Cherenkov light at post-generator level. In this work, EAS simulations at generator level were performed with the \gls{CORSIKA} v7.500 code, using \qgs \cite{QGSJet} and \GHEISHA\cite{Gheisha} packages for the simulation of hadronic processes and EGS4\cite{EGS4} code for the simulation of the EAS electromagnetic component.  The particle production threshold energies in CORSIKA were set to 0.3, 0.3, 0.03, 0.03 GeV, for hadrons, electrons, muons and photons respectively. The thinning option was used with the parameters $10^{-6}$ for the fraction of the primary energy $E$, from which the option is applied, and a maximum weight factor of $10^{-6}\times \left(E/\text{GeV}\right)$. Showers were generated with energies from $10^{14}$ to $10^{19}$ eV (for each decade) and $0^\circ$ inclination. At $10^{18}$ eV, showers were produced with inclinations from $0^\circ$ to $60^\circ$ (in steps of $10^\circ$). For each sample, 50 showers were generated.\\
In order to save information from CORSIKA (a FORTRAN code), a C++ interface was developed to save the generator level data in ROOT\cite{Root} file structures (figure \ref{fig: SkiBin a}). Since it is not feasible to follow and record all particles in ultra high energy showers due to computational and storage costs, the framework is based on saving particle distributions relevant for light emission inside chosen bins "in the sky", \textit{the SkyBins}, which are volumes surrounding the shower axis.  
In the BinTheSky framework, a wavelength range of Cherenkov radiation between 300 - 600 nm was considered, which corresponds to a typical range of the Cherenkov and Fluorescence light detectors \cite{TunkaExp,CTA,YakutskWave}. 

\subsection{BinTheSky framework}
\label{subsection:CORSIKA}

\begin{figure}[!hb]
\begin{center}\centering
       \begin{subfigure}[h]{0.49\textwidth}\centering
                \includegraphics[width=1\textwidth]{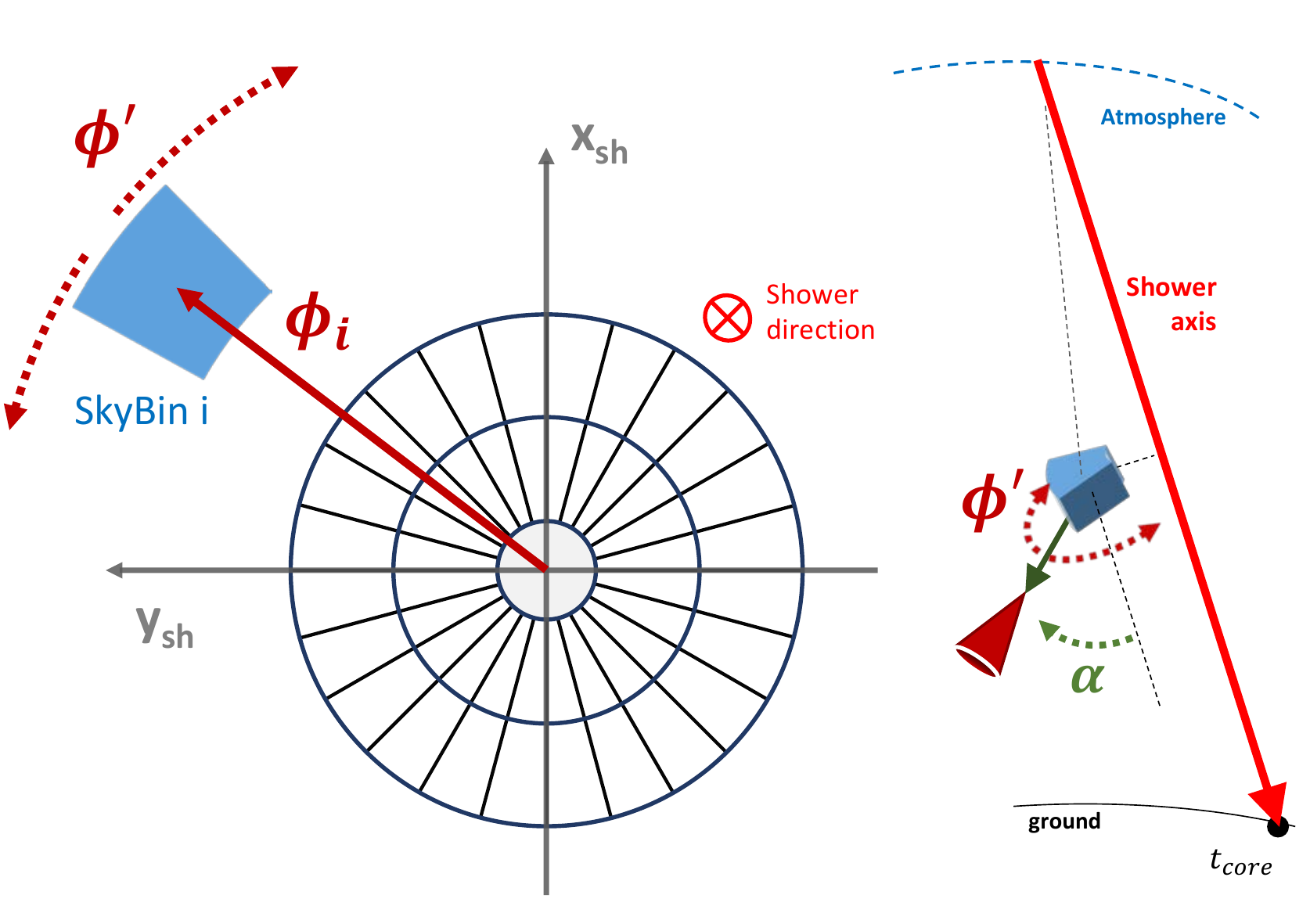}
                \caption{SkyBin Lateral and top view}
                \label{fig: Ch_SkyBin a}
        \end{subfigure}%
                \hspace*{0.00\textwidth}
       \begin{subfigure}[h]{0.49\textwidth}\centering
                \includegraphics[width=1\textwidth]{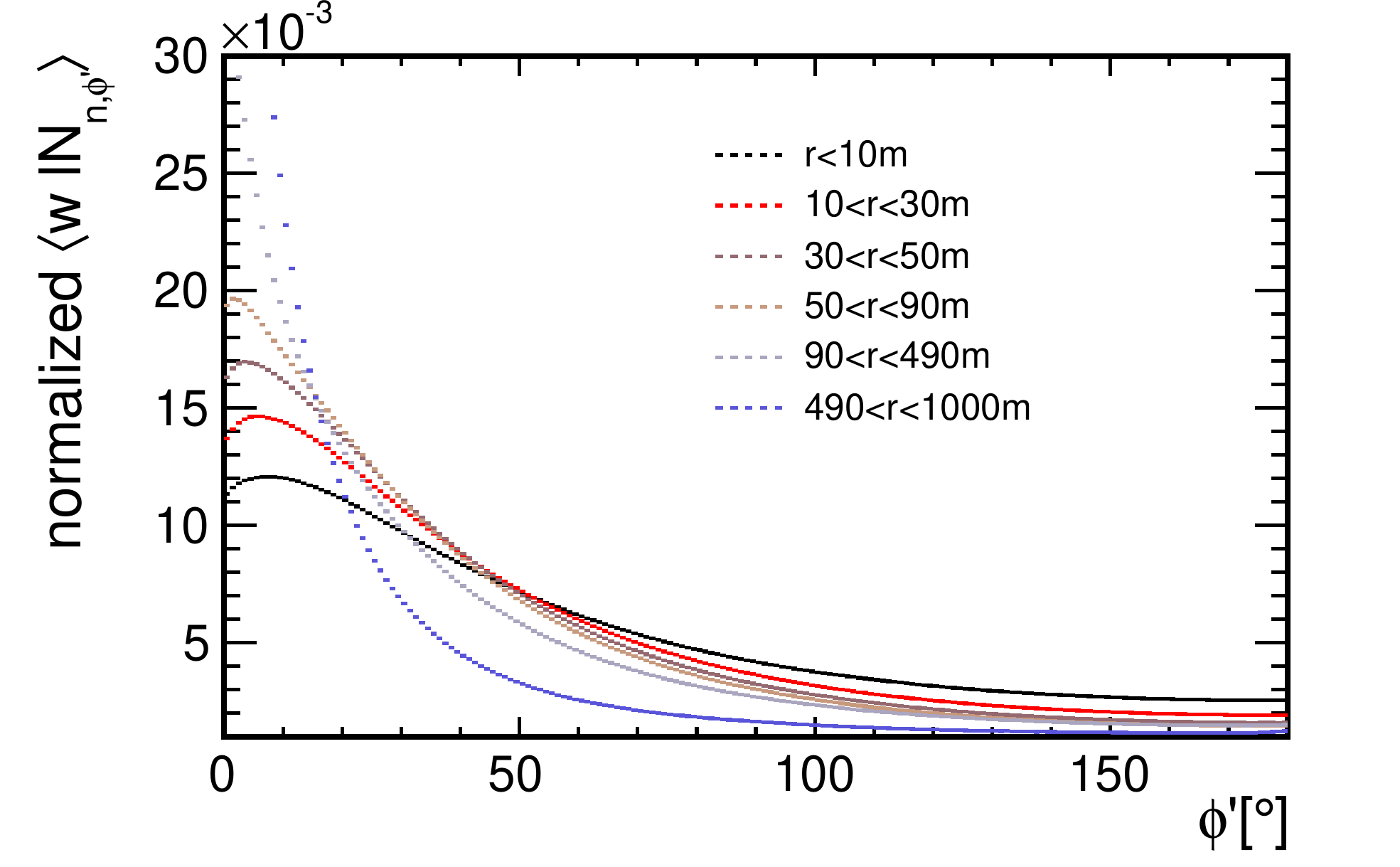}
                \caption{Normalized $\langle wlN_{n,\phi'}\rangle$}
                \label{fig: Ch_SkyBin b}
        \end{subfigure}\\
                
       \begin{subfigure}[h]{0.49\textwidth}\centering
                \includegraphics[width=1\textwidth]{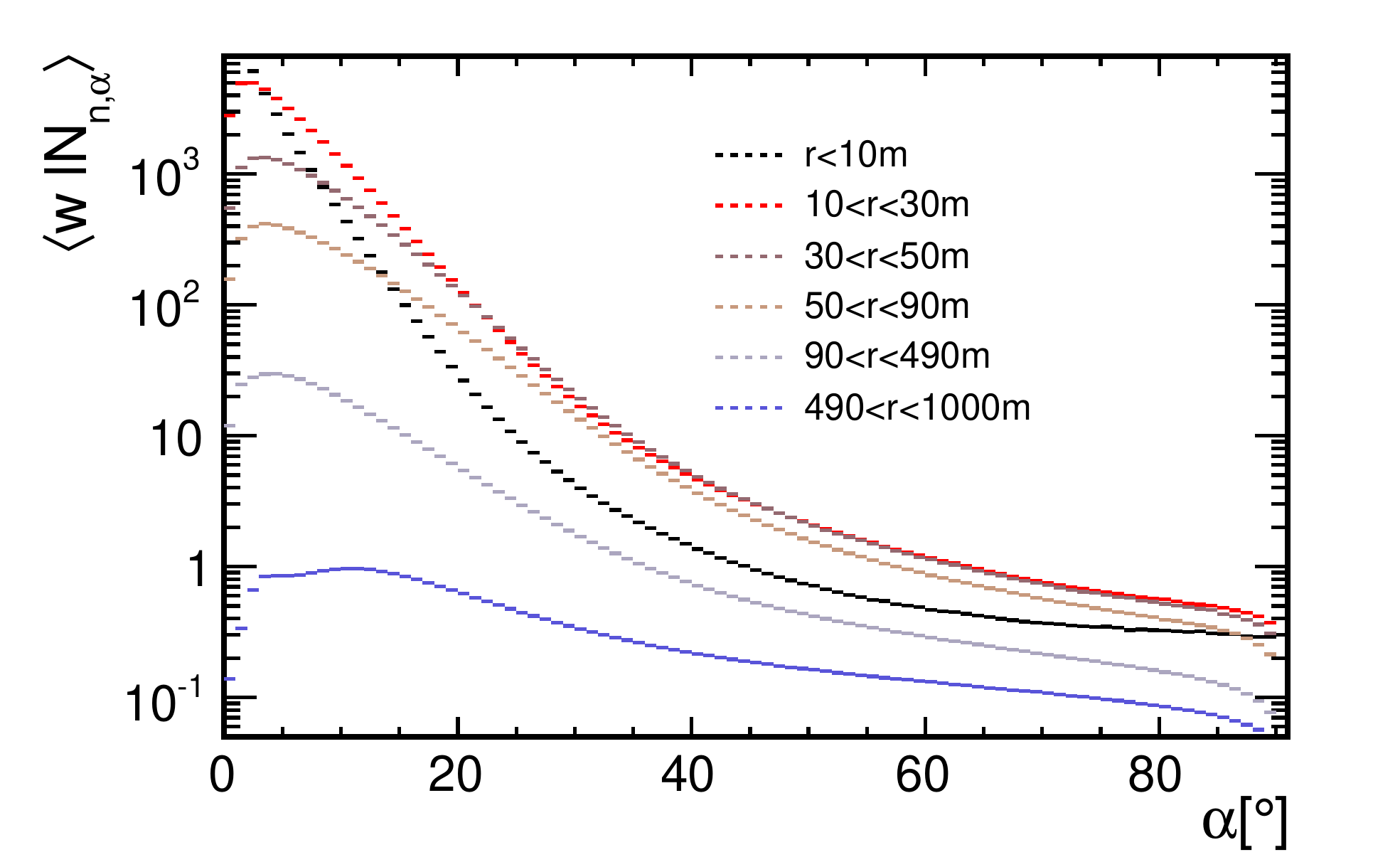}
                \caption{Normalized $\langle wlN_{n,\alpha}\rangle$}
                \label{fig: Ch_SkyBin c}
        \end{subfigure}%
               \begin{subfigure}[h]{0.49\textwidth}\centering
                \includegraphics[width=1\textwidth]{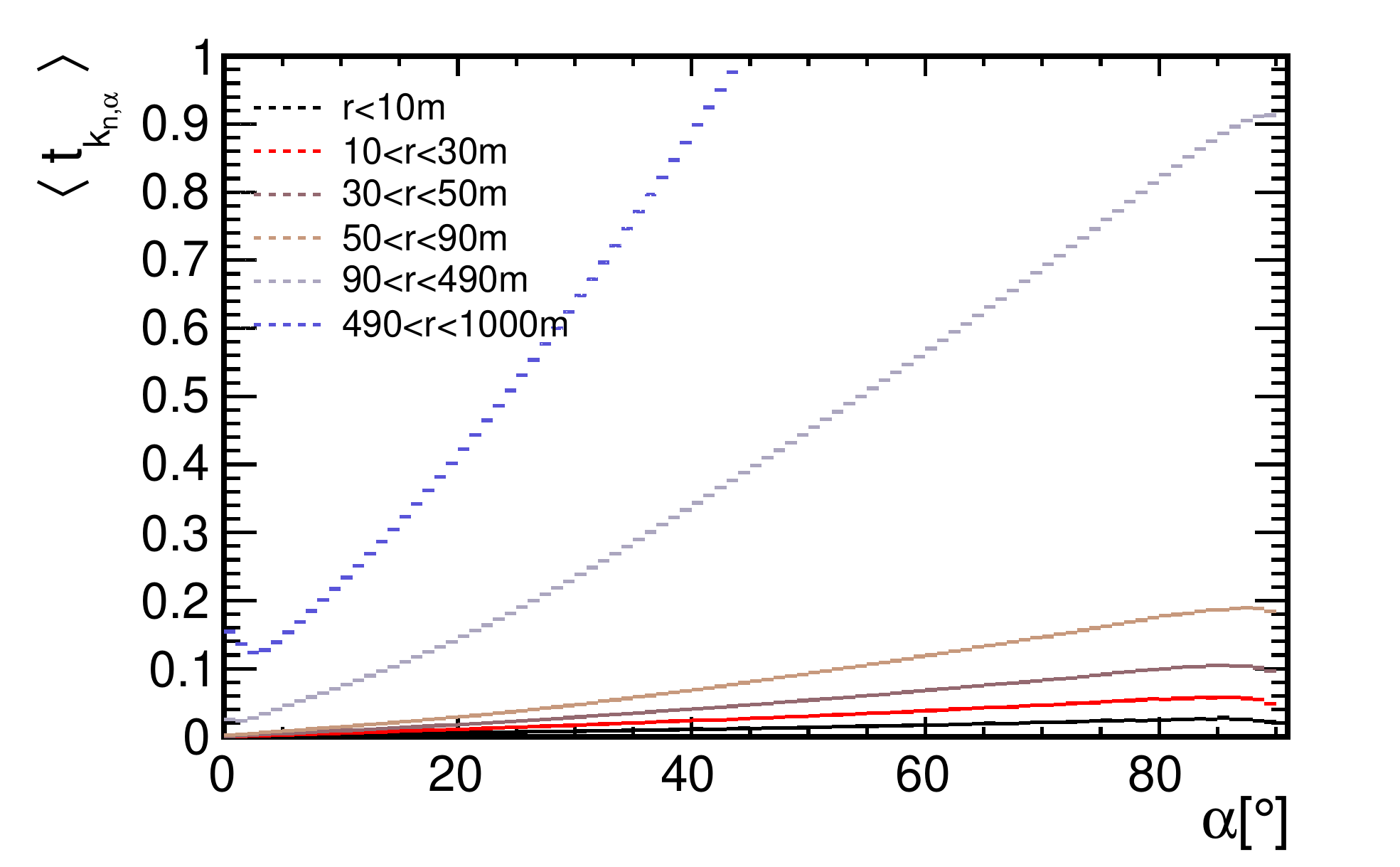}
                \caption{$\langle t_{k_{n,\alpha}}\rangle$}
                \label{fig: Ch_SkyBin d}
        \end{subfigure}\\

\begin{subfigure}[h]{0.49\textwidth}\centering
                \includegraphics[width=1\textwidth]{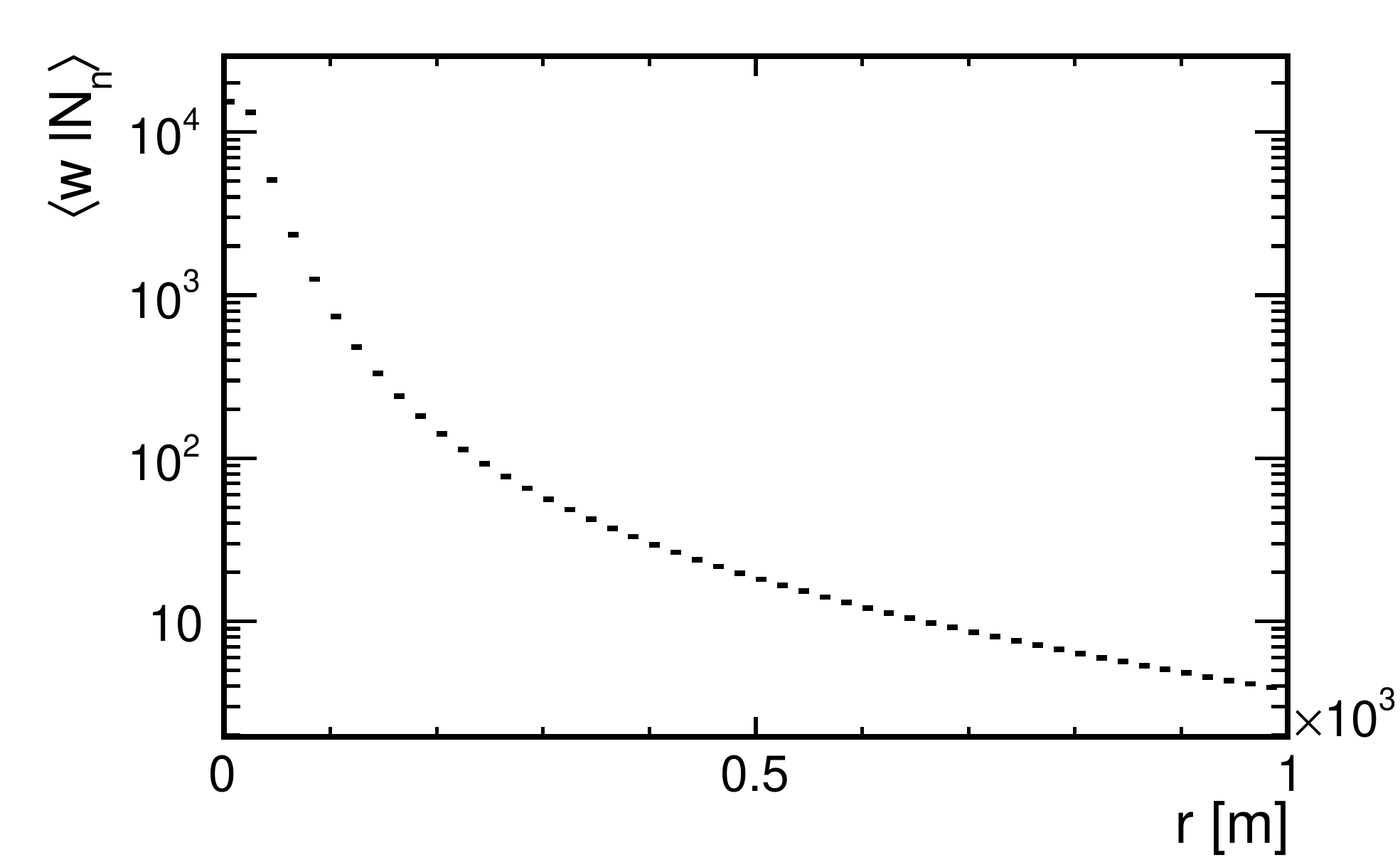}
                \caption{total $\langle wlN_{n}\rangle$ per SkyBin}
                \label{fig: Ch_SkyBin e}
        \end{subfigure}%
               \begin{subfigure}[h]{0.49\textwidth}\centering
                \includegraphics[width=1\textwidth]{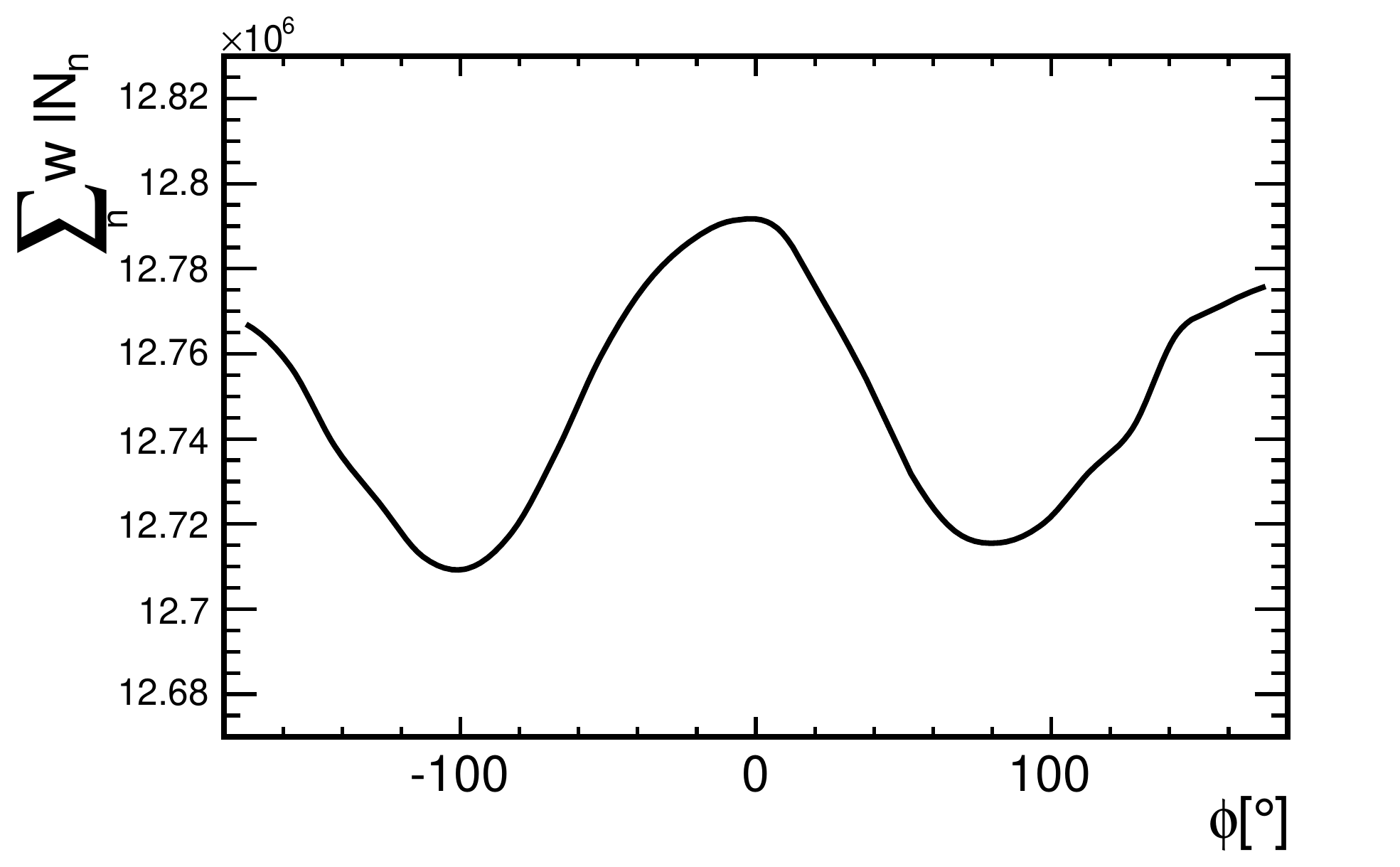}
                \caption{$\sum w lN_{n}$ vs SkyBin $\phi$ }
                \label{fig: Ch_SkyBin f}
        \end{subfigure}%
\caption[]{a) Angular definitions, $\alpha$ and $\phi'$, inside the SkyBin, in a lateral and top view. Normalized $\langle w lN_{n,\phi'}\rangle$ (b) and $\langle w lN_{n,\alpha}\rangle$ (c), for all particles, as function of $\phi'$ and $\alpha$ directions inside the SkyBin, respectively. d) Average time delay $t_k$ as function of $\alpha$. Average $\langle w lN_{n}\rangle$ (e) inside a box (summing all particles directions as function of the distance $r$ to shower axis. Total sum $\sum w lN_{n}$ versus SkyBin $\phi$ direction in a shower. These distribution correspond to one vertical event with $10^{18}$ eV.
}
\label{fig: Ch_SkyBin}
\end{center}
\end{figure}

The \textit{Skybins} are defined with respect to the shower axis - which is the axis defined by the primary particle direction (see figure \ref{fig: SkiBin b} and \ref{fig: SkiBin c}) - with an underlying cylindrical geometry. 
Each SkyBin is a $(\Delta r,\Delta\phi, \Delta z)$ interval with central coordinates $(r_{c},\phi_{c},z_{c})$, where $r$ is the radial distance from a given position in the sky to the shower axis, $\phi$ is the azimuthal angle corresponding to that position and $z$ is its height along the shower axis, from the ground.
As can be seen in figure \ref{fig: Ch_SkyBin} the "sky" around the shower axis is divided in layers in the $z$ axis and several concentric rings of SkyBins, with $\Delta\phi$ and $\Delta r$ segmentation (except for the inner ring bins whose radius is $r=\Delta r/2$). The bins located near the shower axis have thus a smaller volume, appropriate for describing the denser inner shower region.
The size of the SkyBins can be chosen before compiling the CORSIKA with C++. For the case presented in this paper, SkyBin intervals were chosen with $\Delta r=20\text{ m}$, $\Delta\phi=15^{\circ}$ and $\Delta z=100\text{ m}$ (can be defined by the user before compilation).
Using $\Delta z = 100\text{ m}$, yields $~12\text{ g/cm}^{2}$ column depth for the Skybins at sea level, where the atmosphere is denser, and lower column depths for higher altitudes, small enough for the purposes of this study. 

To characterize the Cherenkov photon emission, it is necessary to save information on the directions of relativistic particles in EAS, their numbers and average path length inside the Skybins. Since  there are millions of particles in EAS it is not feasible to save all generated particles, so the distributions of the particle directions inside the SkyBin will be used, instead of the characteristics of each particle.

The directional information of the shower particles is recorded with the angle $\alpha$ between the direction of the shower axis and that of the charged particles (see figure \ref{fig: Ch_SkyBin a}), and independently with $\phi'$ angle, the azimuthal direction of the charged particle with respect to the centre of each SkyBin (see figure \ref{fig: Ch_SkyBin a}). Both variables were saved in intervals $\Delta\alpha=1^\circ$ and $\Delta\phi'=1^\circ$ (user defined before compiling CORSIKA). The path length distribution is saved in the form of the weight
\begin{equation}
L_{i,j}=\left(\sum_n l_n \cdot N^{ch}_{Ckov, n}\cdot w^{Ckov}_{n}\right)_{i,j} \:,
\label{eq: Length}
\end{equation}
for each SkyBin $i$, in the direction $j=\alpha$ or $j=\phi'$ independently (figure \ref{fig: Ch_SkyBin b} and \ref{fig: Ch_SkyBin c}), where $N^{ch}_{Ckov, n}$ is the number of charged particles producing Cherenkov and $l_n$ the length travelled by those particles. 
Only the particles above the Cherenkov energy threshold, fulfilling the condition $v>c/n$ are considered (where $v$ is the particle velocity and $n$ is the air refractive index at the particle position).
 Nevertheless, some particles will have energies close to the threshold, which means they will produce less photons. In this way, we add the weight $w^{Ckov}_{n}=1-\frac{1}{n^2\beta_{n}^2}$, which would appear in the Cherenkov calculations, eq. \ref{eq4:Ck_ph}. We could use another approximation, considering $\beta\sim1$ above the threshold and implement the factor $1-\frac{1}{n^2}$ inside the 3D simulation. Here it is not necessary, but for the sake of completeness, the difference will be briefly discussed in section \ref{section:Corsika}.

All particles are summed up, for $L_{i,j}$, inside each SkyBin as function of $\phi'$ and $\alpha$ and the distributions are saved.
 In the case of the $L_{i,j}$ vs $\phi'$ distribution, the data are folded with respect to $\phi'=0$ within the range $[0^\circ,180^\circ]$, assuming that emissions inside each Skybin to be symmetric with respect to the radial axis.  Additionally, the time delay of Cherenkov photons with respect to a plane shower front propagating at light velocity is also recorded for each Skybin (figure \ref{fig: Ch_SkyBin d}). In this case, the average time delay is saved only as function of $\alpha$ direction, since it less dependent on $\phi'$.
	In figure \ref{fig: Ch_SkyBin e} and \ref{fig: Ch_SkyBin f}, the values for $L_{i,j}$ are shown as function of the distance to the shower axis ($r$) and as function of the shower $\phi$ direction.

\subsection{Light emission and propagation}
\label{subsection:Light}
A particle travelling at a velocity ($v$) larger than the speed of light ($c$) in a medium (with refractive index $n$), emits electromagnetic waves which interfere constructively, giving origin to the so-called Cherenkov radiation.
Cherenkov radiation is emitted in a light cone with an aperture of $\cos\theta = \frac{ct/n}{vt}=\frac{1}{\beta n}$. For ultra-relativistic particles, $\beta\approx 1$ and $\cos\theta_{max}=1/n$. At sea level the refractive index of air is $n=1.00029$\cite{Boley64} and $\cos\theta_{max}= 1.3^\circ$, so the direction of the radiated light is almost coincident with that of the parent particle. The Cherenkov emission energy threshold can be calculated for different shower particles. For electrons  $E_{thr}$ is $21$ MeV, for $\pi$ mesons it is $4.4\cdot10^3$ MeV and for protons $39\cdot10^3$ MeV. The value of $E_{thr}$ for electrons is the lowest, and  since the number of electrons is EAS can be around $95\%$ of the total number of particles in the shower, it can be concluded that Cherenkov radiation in EAS is mainly produced by electrons.
The light is produced in a random position inside the SkyBin $(\Delta r,\Delta\phi, \Delta z)$ and pointing a random direction within the respective ($\Delta\alpha,\Delta\phi'$). Moreover, the light is randomly emitted in the Cherenkov cone.

The number of photons emitted per unit length (travelled by the charged particles), in the wavelength interval $\left[\lambda_1,\lambda_2\right]$, is given by the following equation:
\begin{equation}
\frac{\text{d}N_\gamma}{\text{d}l}=2\pi\alpha\int_{\lambda_1}^{\lambda_2}\left(1-\frac{1}{\beta^2n^2(\lambda)}\right)\frac{1}{\lambda^2}\text{d}\lambda  \:,
\label{eq:CherenkovEmission}
\end{equation}
where $\alpha$ is the fine structure constant and $n$ the refractive index of the medium (see \cite{Nerling2006}).

The distributions shown in figures \ref{fig: Ch_SkyBin} enable a complete characterisation of Cherenkov light production in EAS. The number of emitted photons depends on the number of charged particles above $E_{thr}$ and on the length travelled by those particles.
So, the first step is to calculate the total weighted travelled length in a specific direction $\alpha$ and $\phi'$, in a Skybin $i$ ($L_{i,\alpha,\phi'}$). Using eq. \ref{eq: Length}, the weight $L_{i,\alpha}$ has to be corrected for the relative contribution in the $\phi'$ direction. Therefore, the value $L_{i,\alpha,\phi'}$ is obtained as:

\begin{equation}
L_{i,\alpha,\phi'}= \left(\sum_{n}l_n \cdot N^{ch}_{Ckov, n} \cdot w_n \right)_{i,\alpha,\phi'}\approx L_{i,\alpha} \cdot\frac{L_{i,\phi'}}{L_{i}} \:,
\label{eq:Liap}
\end{equation}
where $L_i$ is the total weight inside the Skybin $i$.
Integrating equation \ref{eq:CherenkovEmission}, the number of photons emitted from a SkyBin, $i$, to some particular direction defined by $(\alpha,\phi')$, with wavelengths in the interval between $\lambda_1$ and $\lambda_2$, can be obtained by:

\begin{equation}
n_{ph,Ckov}^{SkyBin, i,\alpha,\phi'} (\lambda_{1},\lambda_{2})= 2\pi \alpha_e L_{i,\alpha,\phi'}\cdot\left(\frac{1}{\lambda_1}-\frac{1}{\lambda_2}\right) \:.
\label{eq4:Ck_ph}
\end{equation}

The term $\left(1-\frac{1}{n^2\beta_{n}^2}\right)$ is already included in the $w_n$ factor. 
After the Cherenkov light production and prior reaching the detector or ground, the photons are subject to scattering in the atmosphere. This effect has to be taken into account in order to characterize the signal arriving at the detectors. Within the typical wavelengths detected for the EAS experiments (roughly 300-600 nm), 
photon attenuation is due to Rayleigh scattering (photons interacting with air molecules), but also to Mie scattering (photons interacting with aerosol particles). 
The light reaching the ground must be corrected by the attenuation from Mie and Rayleigh scattering. Then, the number of photons arriving to the ground is:
\begin{equation}
n_{ph At ground}^{SkyBin, i,\alpha,\phi'} (\lambda_{1},\lambda_{2}) = n_{ph,Ckov}^{SkyBin, i,\alpha,\phi'} (\lambda_{1},\lambda_{2})\cdot T_{R}\cdot T_{M} \:,
      \label{eq:attenuation}
\end{equation}
where the Rayleigh ($T_R$) and Mie ($T_M$) transmission factors are described in the following sections.

\subsubsection{Rayleigh scattering}
Rayleigh scattering is the elastic scattering of light on particles much smaller than its wavelength. The Rayleigh scattering cross-section has a strong dependence on $\lambda$ ($1/\lambda^4$), and, the total cross-section per molecule of air is given by eq. \ref{eq:Ray} \cite{bucholtz,bookMcCartney}, where $n_s$ is the refractive index for standard air at a given wavelength, $N_s$ the molecular number density ($2.54743\cdot 10^{19}\text{ cm}^{-3}$), and $\rho_n$ is the depolarization factor\footnote{$\rho_n$ accounts for the anisotropy of air molecules. Point-like scatterers should have $\rho_n=0$. It is dependent on the wavelength, but in air $\rho_n$ is expected to be around 0.03, which will be used here.}, that accounts for the anisotropy of the molecules. The depolarization factor has a variance around $60\%$ from the near IR to the UV spectral region, corresponding to a variation with wavelength of approximately 3\% in the Rayleigh-scattering coefficients.

\begin{equation}
\sigma(\lambda)=\frac{24\pi^3\left(n_s^2-1\right)^2}{\lambda^4N_s^2\left(n_s^2+2\right)^2}  \left(\frac{6+3\rho_n}{6-7\rho_n}\right) \:.
\label{eq:Ray}
\end{equation}

The number of photons undergoing Rayleigh scattering, while crossing an atmospheric thickness $\text{d}l$ is given approximately by (see \cite{bucholtz}):
\begin{equation}
     \frac{\text{d}N_{\gamma}}{\text{d}l}=-\frac{\rho_{air}\: k_{A}}{M_{air}}\sigma(\lambda) N_{\gamma} \: ,
\end{equation}
with the Avogrado constant, $k_A$, air density $\rho_{air}$ and molecular mass $M_{air}=28.97\text{ g/mol}$. The transmission coefficient between a depth $X_1$ and $X_2$\footnote{$X$ is the depth in $\text{g/cm}^2$ of crossed atmosphere}, in a trajectory with inclination $\theta$ is then given by:

\begin{equation}
     T_{R}=e^{ -\frac{\vert X_{1}-X_{2}\vert}{\cos\theta} \sigma(\lambda) k_A/M_{air} }   \:.
     \label{eq:TRay}
\end{equation}

\subsubsection{Mie scattering}
Mie scattering is the scattering of photons by particles whose size is comparable to their wavelength. These particles are aerosols (dust, pollutants, liquid droplets) with typical radius around 0.1 to 10 $\mu$m. The aerosols dimensions, chemical composition and even their shape contribute to the Mie scattering cross-section. The aerosol density in the atmosphere can be considered as falling exponentially with altitude so the Mie scattering is estimated, based upon the model of Elterman \cite{bookElterman}, by:
\begin{equation}
      \frac{\text{d}N_{\gamma}}{\text{d}l}= -\frac{N_{\gamma}}{l_{M}}e^{-h/h_{M}}  \:,
      \label{eq:Mie}
      \end{equation}
where $h$ is the height, $h_{M}$ is the scale height factor and $l_{M}$ is the Mie scattering mean free path, which depend on the atmospheric properties. Values of $h_{M}=1200$ m and $l_{M}=14000$ m (a typical Mie scattering mean free path at $\sim400$ nm)\cite{bookSokolsky}, were used in the calculations. 

Using $\text{d}l = 
\text{d}h / \cos \theta$ ($\theta$ being the angle between the vertical and the photon path), the Mie scattering transmission coefficient is given by:
\begin{equation}
      T_{M}=e^{-\tau(\lambda) \frac{h_{M}}{l_{M}(\lambda)\cos \theta} \left[ e^{-h_{1}/h_{M}}-e^{-h_{2}/h_{M}} \right] }  \:.
      \label{eq:TMie}
\end{equation}

It should be noted that $h_{M}$ and $ l_{M}$ strongly depend on the composition of the atmosphere and therefore accurate measurements of the atmospheric aerosol composition and corresponding scale hight are fundamental for the characterisation of Mie scattering. Furthermore, the parameter $\tau(\lambda)$ was added for the wavelength dependence. According to \cite{MieWavelength}, it gives $\tau(\lambda)=\left(\frac{\lambda}{355\text{ nm}}\right)^{0.767}$.

\subsection{Time calculations}
\label{subsection:Time}

The time calculations for Cherenkov photons arriving at the ground are computed as time delays with respect to a plane shower front, travelling at the velocity of light and arriving on the ground at the same point as the Cherenkov photons.
The final photon time delay, $t_{T}$, at some position on the ground is given by: 
 
\begin{equation}
t_{T}=t_g + t_k + t_{n} \:,
\label{eq4: Timing}
\end{equation}

where $t_g$ is the geometric time delay, $t_k$ and $t_{n}$ are the the kinematic and refractive index time delays.

Photons are generated inside the Skybin, at some random emission position $x_{re}=(r_{sh},\phi_{sh},z_{sh})$ in the shower frame of reference. Then, they travel along their final direction reaching the ground at a position $x_{G}=(r_{sh,G},\phi_{sh,G},z_{sh,G})$ in the shower frame. Considering a plane shower front travelling at the speed of light along the $z_{sh}$ direction, the photon geometric time delay is given by:
\begin{equation}
ct_g = \| \vec{x_{re}}-\vec{x_{G}}\|  - \left(z_{sh}-z_{sh,G}\right) \:.
\end{equation}

Cherenkov photons in air are slower than $c$ and the corresponding time delay, $t_n$, related to the air refractive index is thus given by:
\begin{equation}
c t_n= \int_{h}^{h_{G}} \left[n(z)-1\right] dz /\cos \theta  \:.
\label{eq: refindex}
\end{equation}

To simplify the use of eq. \ref{eq: refindex}, and accelerate the running time,  a simple atmospheric layer was considered with the depth given by $X(h)=1117.9 \:e^{-h/6839}\text{ g/cm}^2$. The refractive index evolves linearly with the depth, so we consider the Lorentz-Lorenz relation $\frac{n^2-1}{n^2+2}\propto\text{const}\cdot\rho$. The atmospheric density $\rho$, can be substituted by the depth $X$ and use the reference values of $X=1033\text{ g/cm}^2$ and $n=1.000292$ at see level. Using this values the depth gives differences lower than $2\text{ g/cm}^2$ for $h>15000$ m and lower than $20\text{ g/cm}^2$ otherwise, with respect to CORSIKA atmosphere definitions. For other purposes this would not be sufficiently accurate.
However, for the final $t_n$ delay, it gives differences lower than $0.2\%$ ($\sim2\cdot10^{-2}$ ns) for $h>15000$ m and lower than $2.5\%$ ($\sim4\cdot10^{-2}$ ns) otherwise. It is sufficient to calculate the refractive index time delay. 

Most of particles producing the Cherenkov light are already delayed with respect to the shower front and this is the origin of the kinematic time delay, $t_k$ term. The average particle time delays in CORSIKA were saved for each SkyBin as function of $\alpha$ (see figure \ref{fig: Ch_SkyBin d}) and used as the kinematic time delay factor for Cherenkov photons in the corresponding $\alpha$ intervals.
In this paper, we will not consider directly the final total time, but the time delay with respect to a shower front plane travelling at the speed of light. In this way, for inclined events, the problem of different times in the ground, on the forward and backward direction of the shower, is solved.

\section{BinTheSky framework validation with CORSIKA}
\label{section:Corsika}

At lower energies, it is still possible to simulate the Cherenkov light with the CORSIKA generator. The results obtained with CORSIKA at $10^{14}$ eV and $10^{15}$ eV will be compared here with the 3D simulation results. Note that, in CORSIKA, only the photons in a square with 1500 m around the shower core were saved and for that we get an output per event with around 15 Gb and 170 Gb at $10^{14}$ eV and $10^{15}$ eV, respectively. The comparison is made for only one event for each shower energy and inclination.

\begin{figure}[h]
\begin{center}\centering
       \begin{subfigure}[h]{0.495\textwidth}\centering
                \includegraphics[width=1\textwidth]{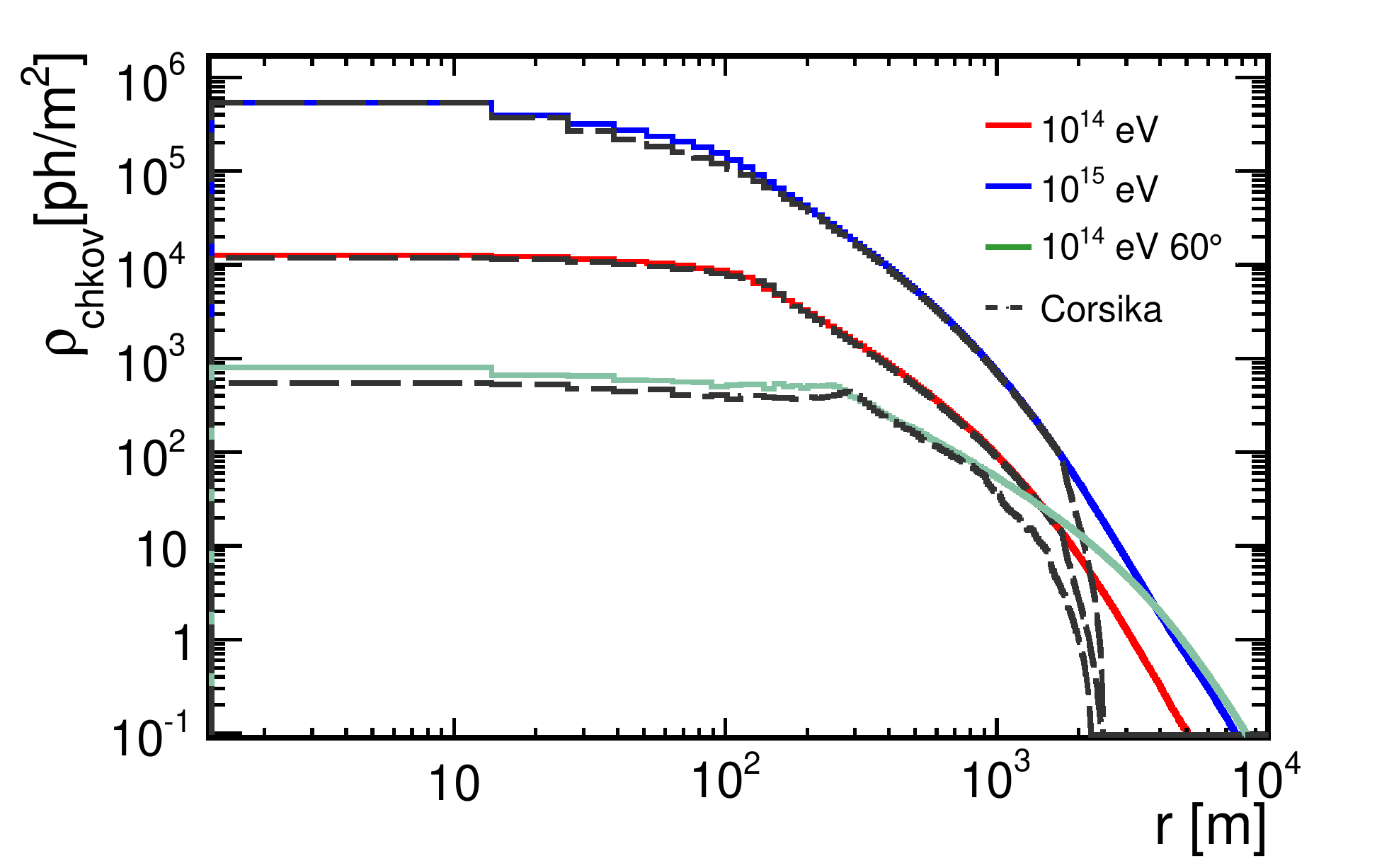}
                \caption{Lateral density of photons.}
                \label{fig: Validation a}
        \end{subfigure}%
       \begin{subfigure}[h]{0.495\textwidth}\centering
                \includegraphics[width=0.909\textwidth]{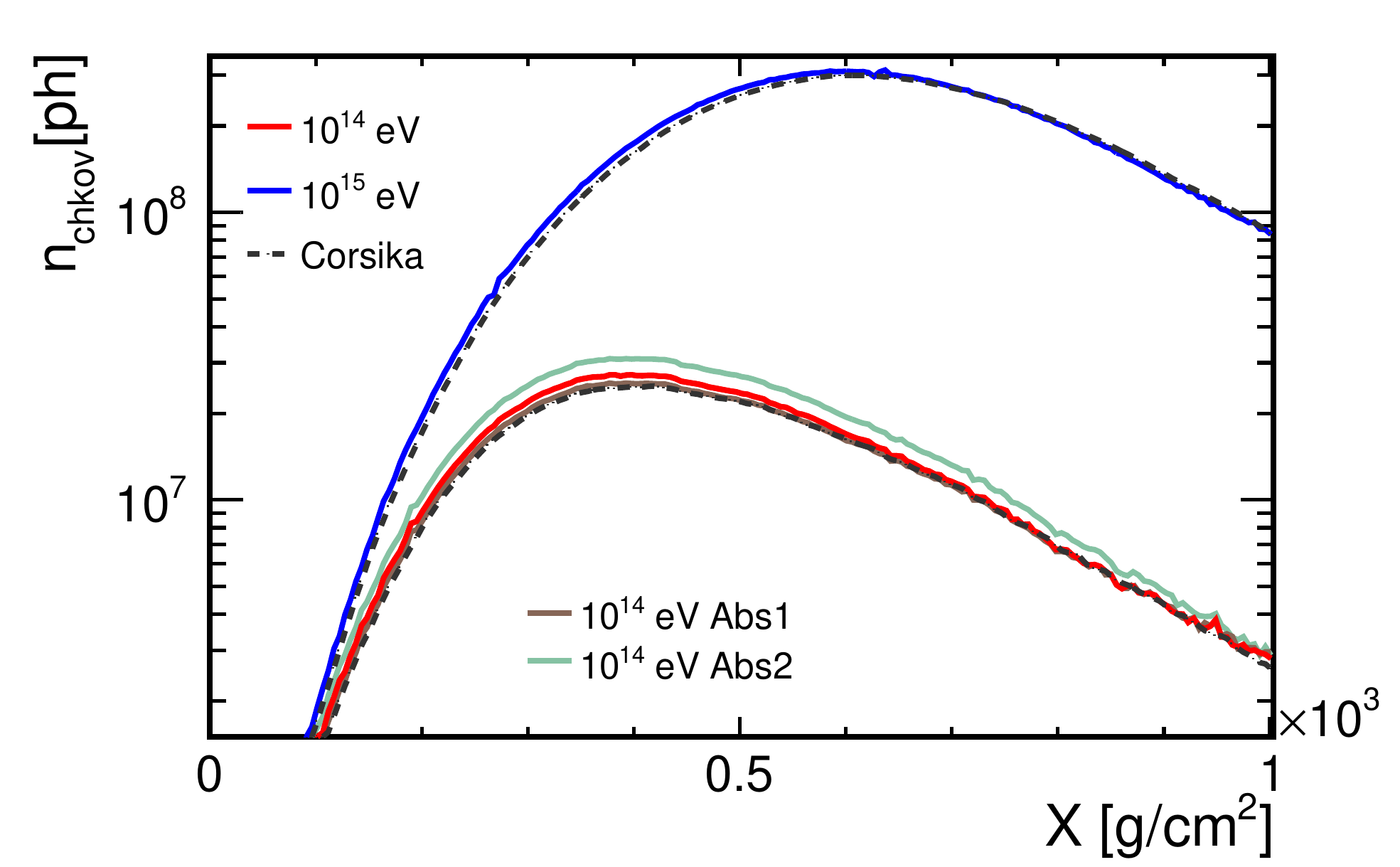}
                \caption{Longitudinal Cherenkov profile}
                \label{fig: Validation b}
        \end{subfigure}%
\\
       \begin{subfigure}[h]{0.495\textwidth} \centering
                \includegraphics[width=1\textwidth]{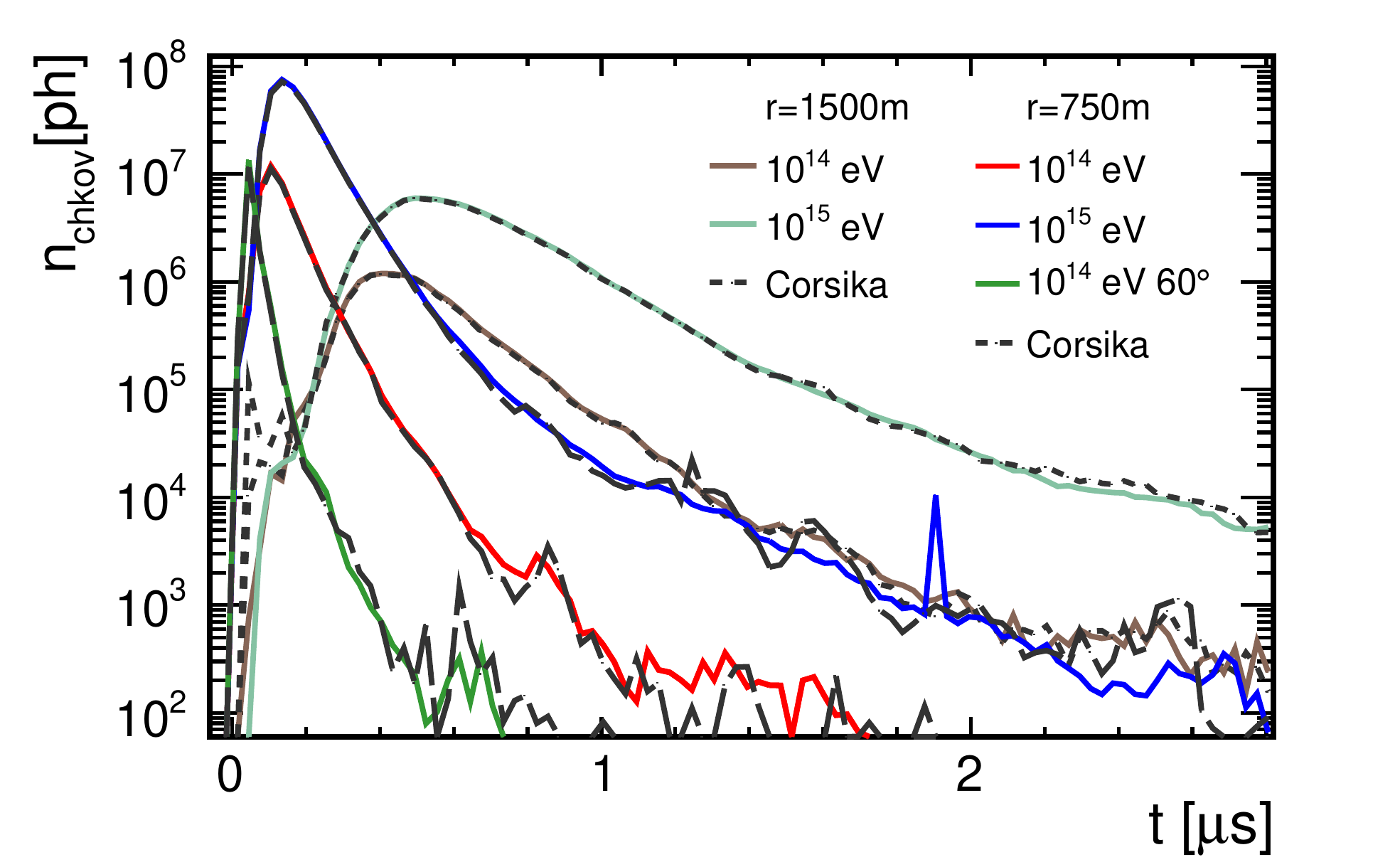}
                \caption{total time delay distributions}
                \label{fig: Validation c}
        \end{subfigure}%
               \begin{subfigure}[h]{0.495\textwidth}\centering
                \includegraphics[width=1\textwidth]{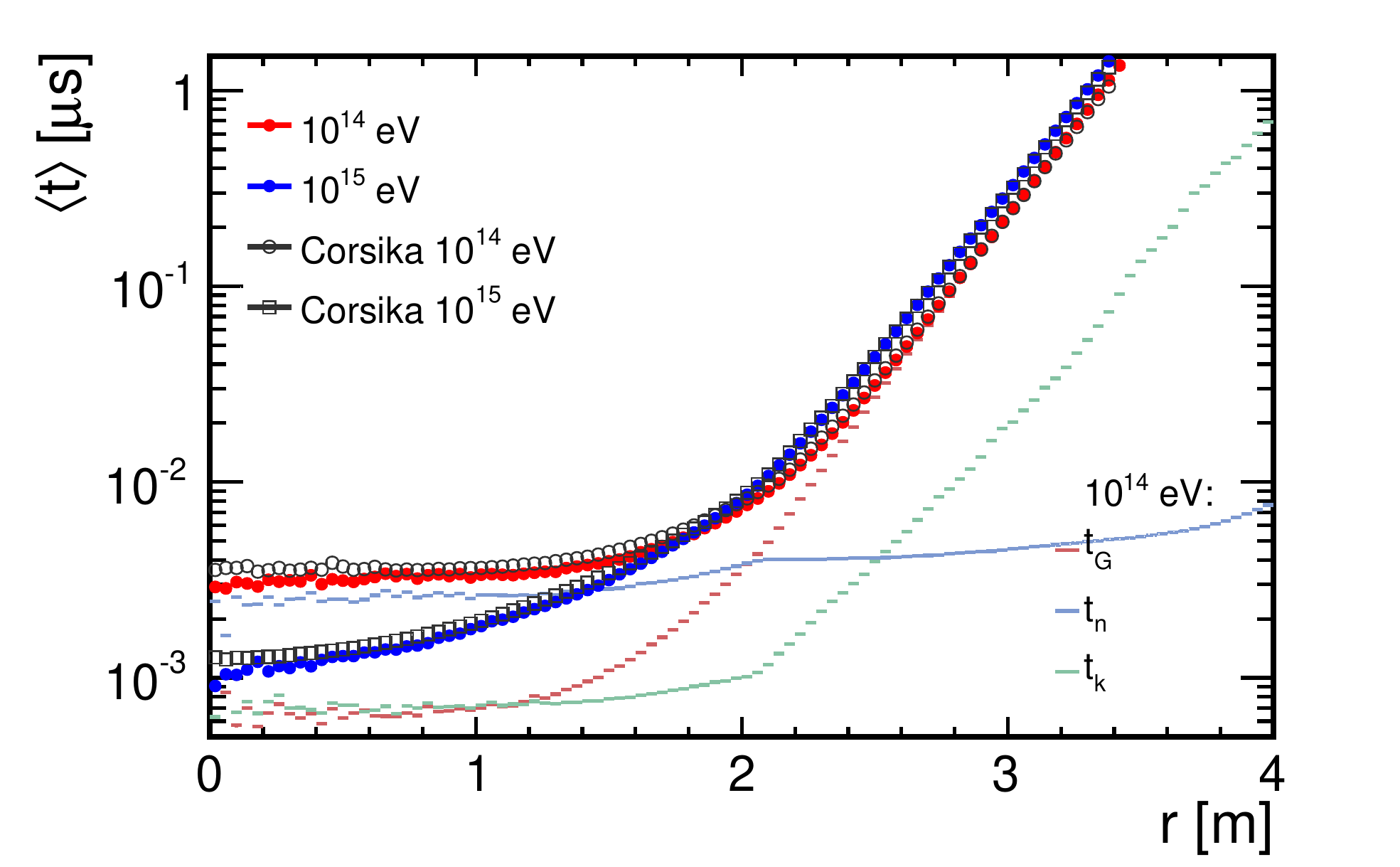}
                \caption{Average time delays}
                \label{fig: Validation d}
        \end{subfigure}%
\caption[]{Results obtained for one event. a) Density of photons as function of the radius to the shower core at a depth of $100\text{ g/cm}^2$, for the 3D simulation at $10^{14}$ eV and $10^{15}$ eV with $0^\circ$, and at $10^{14}$ eV  with $60^\circ$, in red, blue and green full lines, compared with the CORSIKA in dashed lines. b) longitudinal profile of Cherenkov photons arriving to the ground, with the same color definitions. It also includes a 3D simulation for the same  $10^{14}$ eV with a cleaner atmosphere (abs2) and CORSIKA atmospheric attenuations (abs1), in green and brown respectively. c) the total time delay of photons at a radius of 750 m and 1500 m (50 m interval) for both energies. d) Average time delays, for the vertical showers, as function of the radius, time components included for the $10^{14}$ eV event.
}
\label{fig: Validation}
\end{center}
\end{figure}

In the figure \ref{fig: Validation a}, the lateral density of photons in the 3D simulation (full line) is compared with the CORSIKA results (dashed line). The results of both codes are similar up to a 1500 m radius. Above 1500 m CORSIKA photons were not recorded, as explained above.\\
The longitudinal profile of Cherenkov photons arriving to the ground is shown in figure \ref{fig: Validation b}. In these distributions a slight difference can be seen between the 3D simulation and CORSIKA. This difference comes from different atmospheric attenuations. In this context, the 3D simulation was repeated for different atmospheric parameter for the same $10^{14}$ eV event considering 3 atmospheric attenuations: in red, the attenuations used in this the paper from eq. \ref{eq:TRay} and \ref{eq:TMie}; in green a cleaner atmosphere was considered with parameters $h_{M}\sim2800$ m and $l_{M}\sim50000$\cite{MUSSAaerosol} for Mie scattering; and in brown total atmospheric attenuations were used from CORSIKA. The 3D simulation gives the same result as the CORSIKA program using the latter attenuations. The cleaner atmosphere will result in around $\sim10\%$ more light, while the attenuations from \ref{eq:TRay} and \ref{eq:TMie} yield $\sim5\%$ more light than the parametrization used in CORSIKA. This illustrates the importance of a good knowledge over the atmosphere for a clear understanding of the signals measured by the cosmic rays detectors. \\

The other important distributions to verify are the time distributions. In figure \ref{fig: Validation c}, the total delay on the arrival time of photons at a radius of 750 m and 1500 m from the shower core is plotted, showing a good agreement with the CORSIKA profiles. The average time delays are shown in the figure \ref{fig: Validation d} and it can be seen that CORSIKA and the 3D simulation have similar results. For higher radius the total time delay is dominated by the geometric delay, while at lower radius the refractive index delay dominates. 

In the figures \ref{fig: Validation a} and \ref{fig: Validation c}, the lateral density and time distributions are shown also for events with $10^{14}$ eV and $60^\circ$, revealing a good agreement between CORSIKA and the 3D simulation result.\\

\begin{figure}[h]
\begin{center}\centering
       \begin{subfigure}[h]{0.495\textwidth} \centering
                \includegraphics[width=1\textwidth]{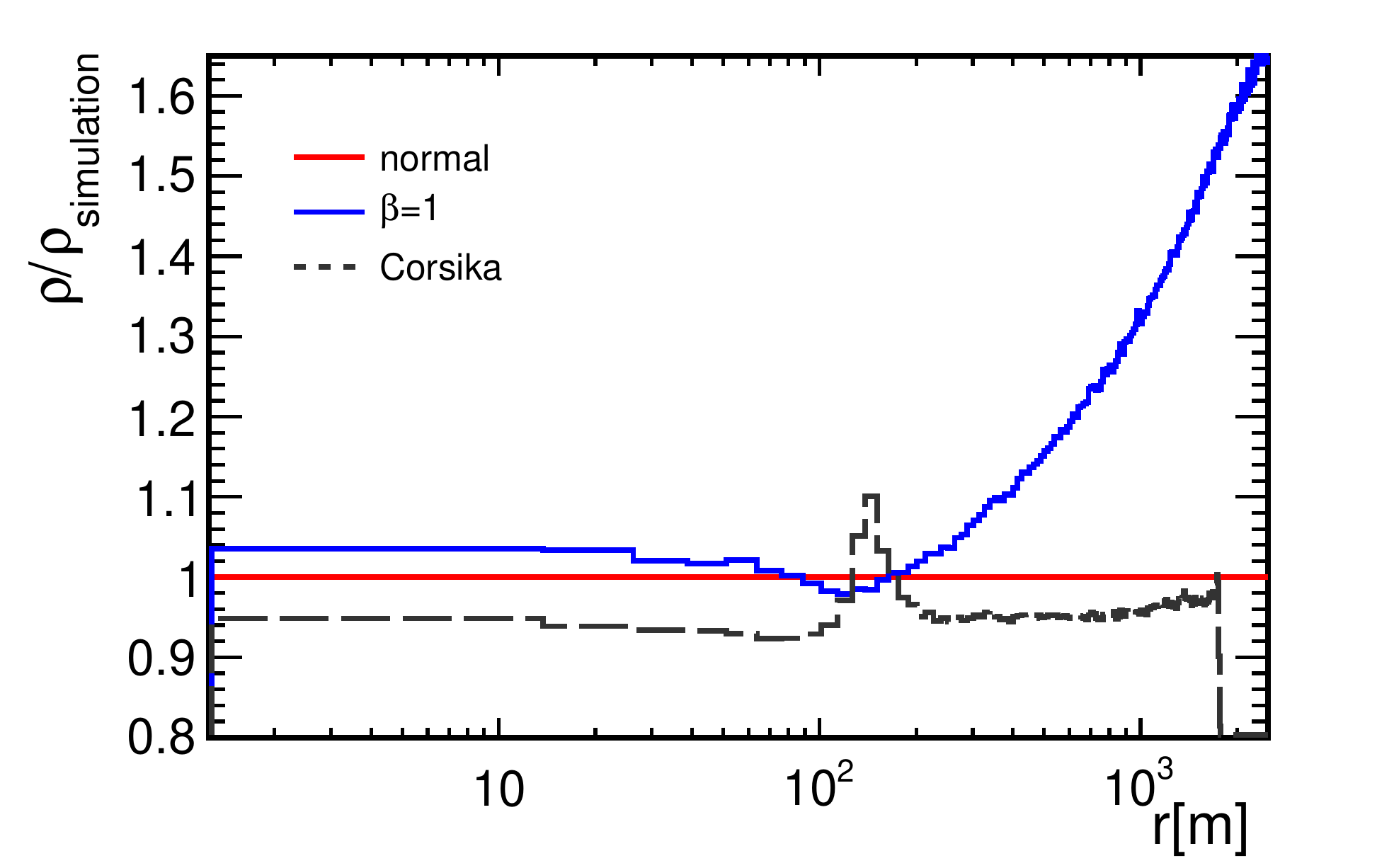}
                \caption{difference in the lateral density of photons}
                \label{fig: Validation2 c}
        \end{subfigure}%
       \begin{subfigure}[h]{0.495\textwidth}\centering
                \includegraphics[width=0.909\textwidth]{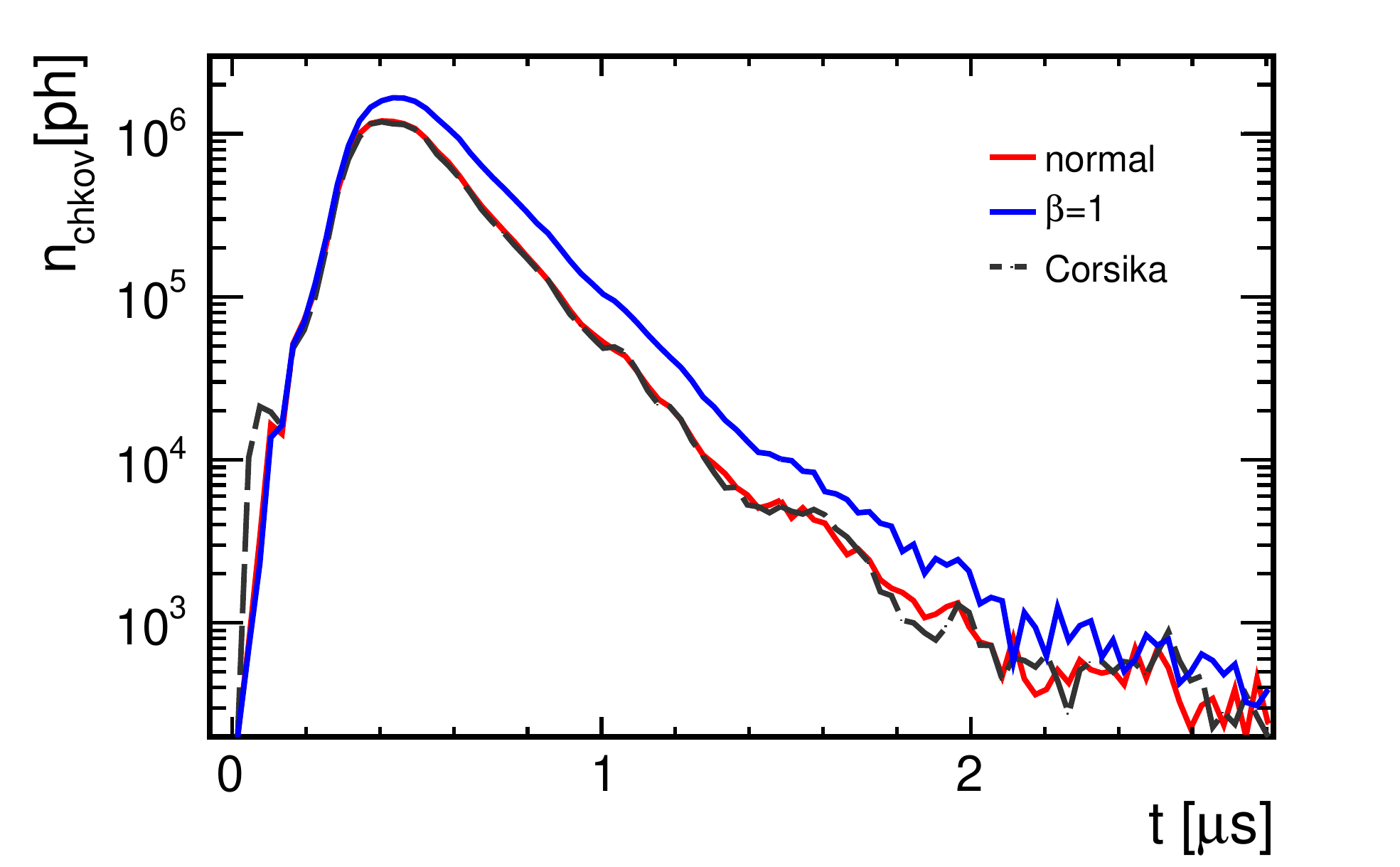}
                \caption{total time delay}
                \label{fig: Validation2 b}
        \end{subfigure}%
\caption[]{Results obtained for one event at $10^{14}$ eV and $0^\circ$, this event was simulated with the 3D simulation, in red, considering the weight $w^{Ckov}_{n}=1-\frac{1}{n^2\beta_{n}^2}$ in eq. \ref{eq: Length}. In blue, the 3D simulation under the approximation of $\beta\sim1$ and in dashed lines the CORSIKA output. The fraction of lateral density of photons, between the simulations and the normal 3D simulation are shown in a). The total time delay are drawn in b).
}
\label{fig: Validation2}
\end{center}
\end{figure}

To simplify the CORSIKA output for the 3D simulation, the approximation $\beta=1$ in eq. \ref{eq: Length} had been considered for particles with energy above the Cherenkov energy threshold and the term $\left(1-\frac{1}{n^2\beta_{n}^2}\right)$ was included in eq. \ref{eq4:Ck_ph} inside the 3D simulation (at generator level, when the travelled length is filled). In such case, the number of Cherenkov photons was overestimated. For future reference we show the effect of this overestimation in figure \ref{fig: Validation2} for the lateral shape and for the time delay distributions. We can see that, using $\beta=1$, as the radius increases, the number of photons increases with respect to the actual value. The particles with larger opening angle with respect to the shower axis have lower energy, so we would overestimate more photons far from the core. Close to the core the difference is around $\sim5\%$ and far from the core around $\sim20\%-60\%$. The peak at $r\sim150$ m in CORSIKA with respect to the 3D simulation, is due to the random in first $\alpha<1^\circ$ bin. 
The time delay distribution will have bigger tales, since we are increasing the weight of the latest particles (which have less energy).\\

\FloatBarrier

\section{Ground distributions}
\label{section:Ground}

\begin{figure}[h]
\begin{center}\centering
       \begin{subfigure}[b]{0.49\textwidth}\centering
                \includegraphics[width=1\textwidth]{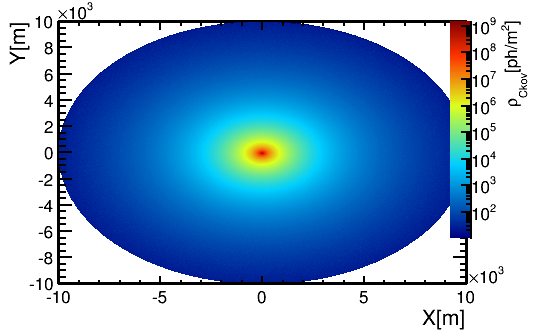}
                \caption{$\theta=0^\circ$}
                \label{fig: ChkovMap a}
        \end{subfigure}%
                \hspace*{0.00\textwidth}
       \begin{subfigure}[b]{0.49\textwidth}\centering
                \includegraphics[width=1\textwidth]{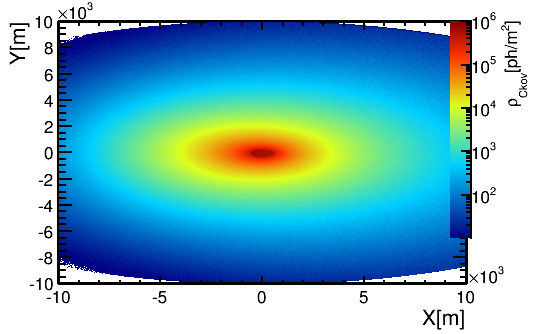}
                \caption{$\theta=60^\circ$}
                \label{fig: ChkovMap b}
        \end{subfigure}
\caption[]{density of Cherenkov light arriving at ground for an event with $10^{18}$ eV and zenith angles $0^\circ$ (0) and $60^\circ$ (b). Shower axis aligned with the y axis in the inclined case. }
\label{fig: ChkovMap}
\end{center}
\end{figure}

The framework defined in the previous section allows one to simulate Cherenkov light without shape parametrizations. The lateral distributions on the ground as well as the time distribution can be obtained. In figure \ref{fig: ChkovMap}, the map shows all Cherenkov light arriving at the ground for an event with $10^{18}$ eV, with zenith angles $0^\circ$ (a) and $60^\circ$ (b).\\

\begin{figure}[h]
\begin{center}\centering
       \includegraphics[width=0.55\textwidth]{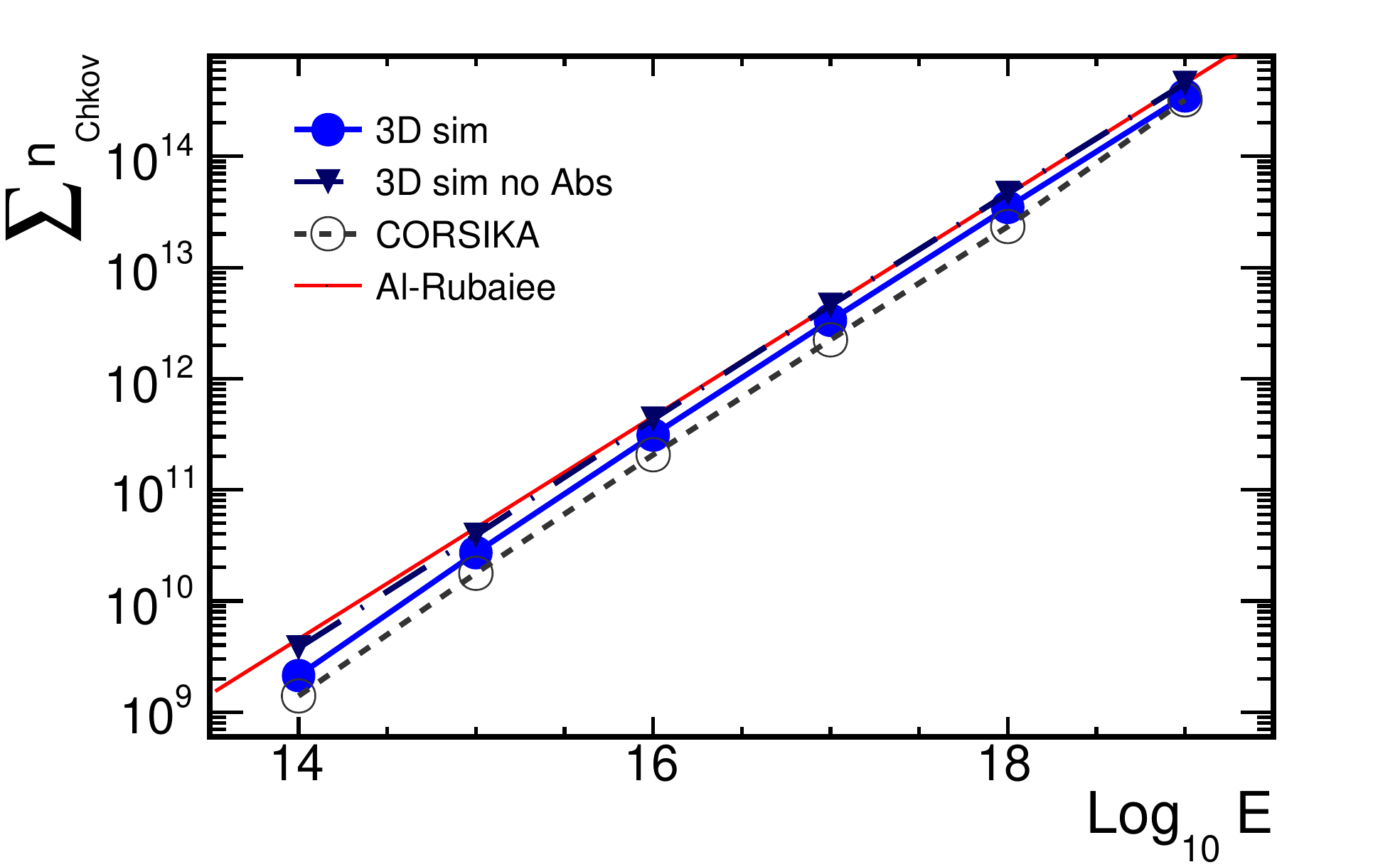}
\caption[]{Total amount of Cherenkov photons for the 3D simulation including (dark triangles) and without attenuations (blue circles), CORSIKA (open circles) and from \cite{Rubaiee2015} (red line).
}
\label{fig: TotPh}
\end{center}
\end{figure}

In \cite{Rubaiee2015}, the total number of Cherenkov photons, neglecting atmospheric absorption, was predicted as $N_{Chkov}=3.7\cdot10^{3}\frac{E_0}{\beta_t}$, where $\beta_t=37.1\text{ g/cm}^2\cdot 2.2\text{ MeV.g/cm}^{2}=81.4$ MeV is the critical energy for the ionization losses. It can be compared with our simulation (with and without atmospheric attenuations) and with CORSIKA in figure \ref{fig: TotPh}. The parametrization agrees with the 3D simulation without attenuation. Again, due to the attenuation, the 3D simulation produces approximately $\sim10\%$ more than CORSIKA. In this case, Cherenkov photons are produced in CORSIKA, but not propagated to the ground, and the attenuations are simply calculated along the shower axis starting at production height.

\FloatBarrier
\subsection{Lateral distributions}
\label{subsection:LDFs}
\begin{figure}[h]
\begin{center}\centering
       \begin{subfigure}[b]{0.49\textwidth}\centering
                \includegraphics[width=1\textwidth]{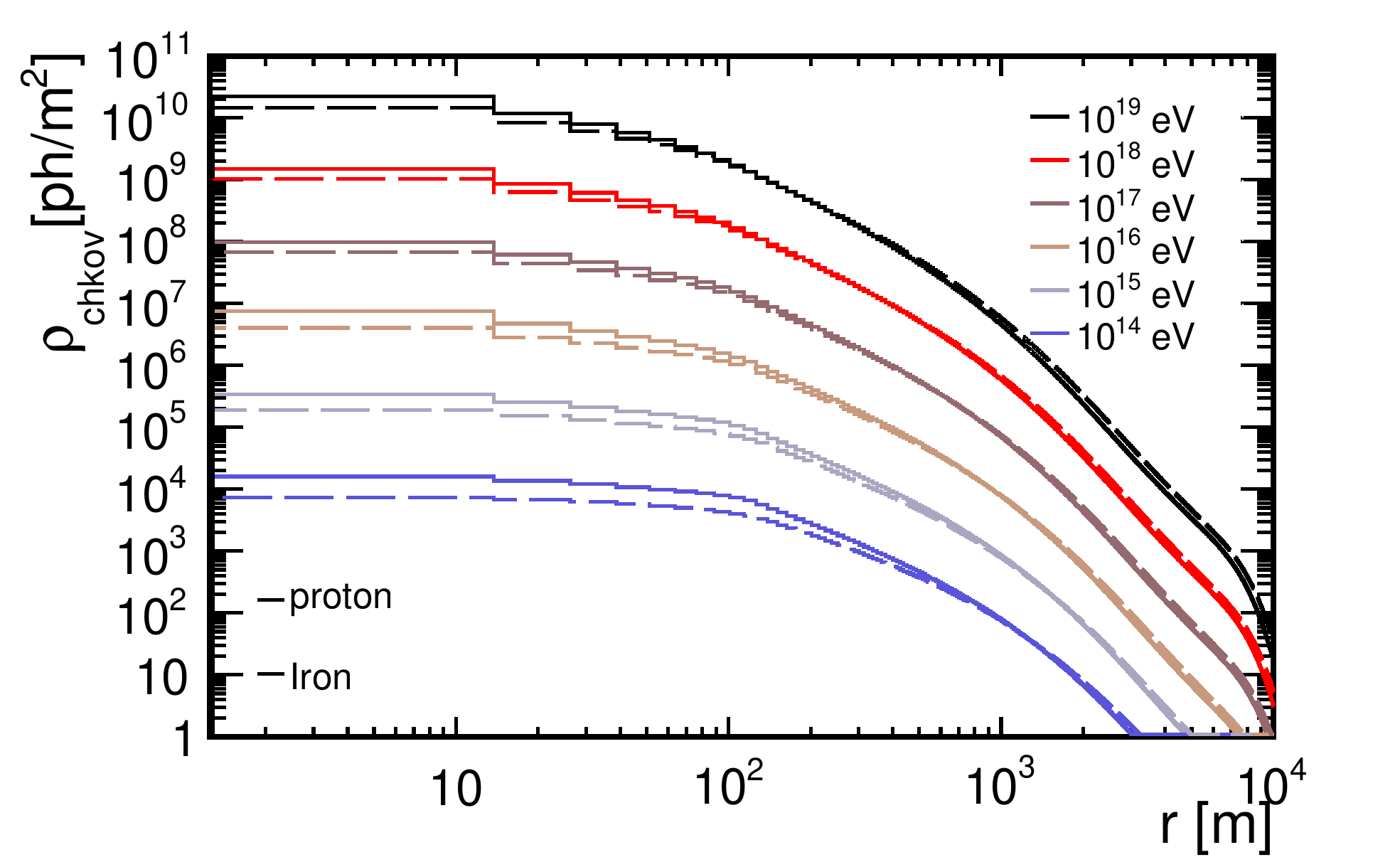}
                \caption{photons for proton and Iron showers}
                \label{fig: LDF a}
        \end{subfigure}%
                \hspace*{0.00\textwidth}
       \begin{subfigure}[b]{0.49\textwidth}\centering
                \includegraphics[width=1\textwidth]{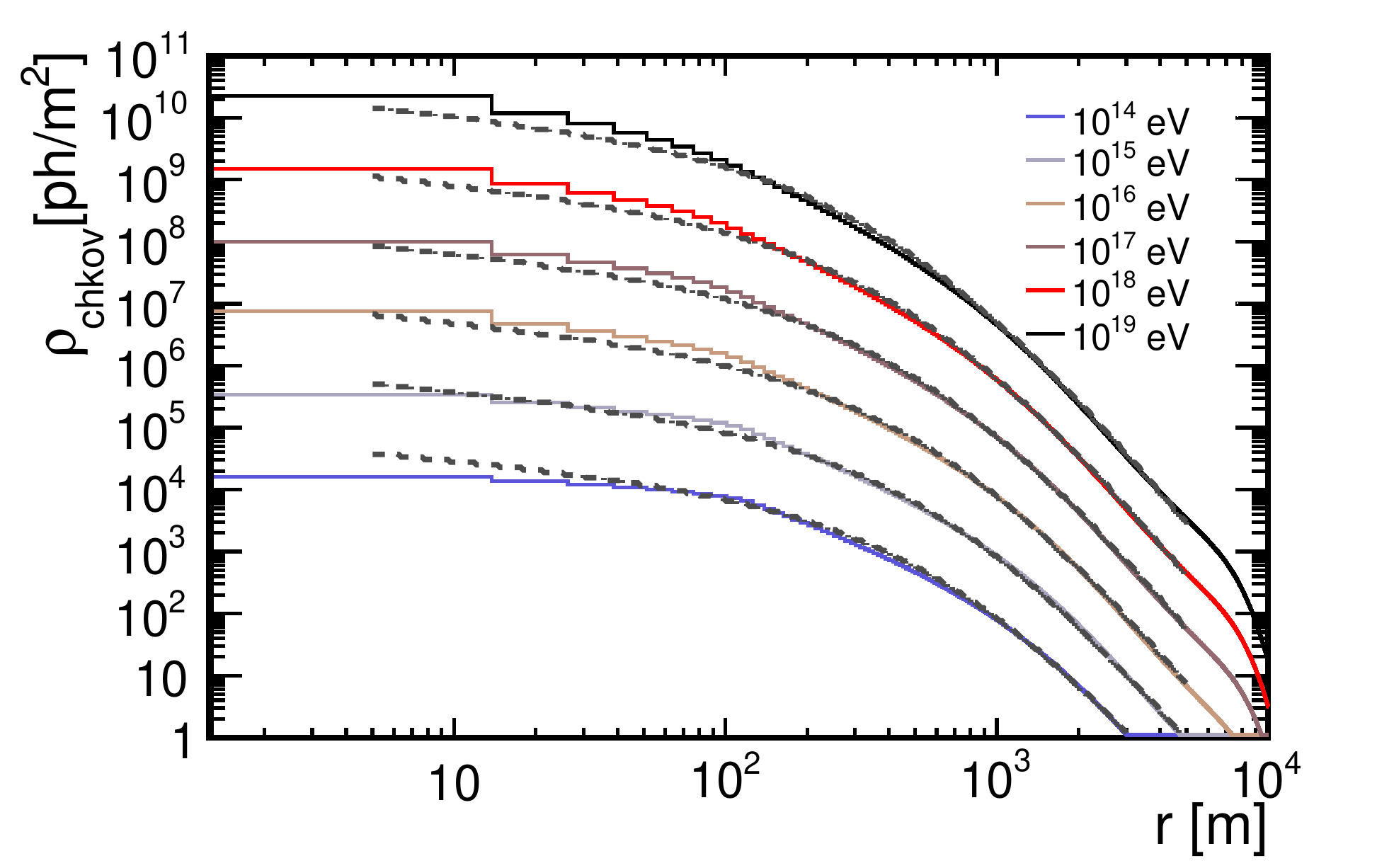}
                \caption{fit for proton showers}
                \label{fig: LDF b}
        \end{subfigure}
\caption[]{a) density of Cherenkov photons as function of the radius $r$  for vertical showers with energies between $10^{14}$ to $10^{19}$ eV, for proton and iron showers. b) Cherenkov photons density for proton showers with respective fit to equation \ref{eq: LDFfit}. 
}
\label{fig: LDF}
\end{center}
\end{figure}

The lateral density of Cherenkov photons can be obtained and parametrized. 
In the case of non-vertical events, the arrival position on the ground does not correspond to the same shower path length, so photons have different attenuations and production points. In this way, we show the lateral density of photons for vertical events with energies between $10^{14}$ to $10^{19}$ eV, for proton and iron showers, in figure \ref{fig: LDF a}.

To parametrize the photon density distribution on the ground, the following equation, inspired by the NKG\cite{Nishimura,GreisenLDF} parametrization, was considered: 
\begin{equation}
\rho(r, E) = \left[	c_0 + c_1\cdot E_s \frac{ }{ } \right]  \left(r\frac{ }{ }\right)^{\lambda_a + \lambda_{a1}\cdot E_s}\left(1+\frac{r}{R_{s}+R_{s1}\cdot E_{s}}\right)^{\lambda_b} \:,
\label{eq: LDFfit}
\end{equation}
where $E_s=\left(\log_{10}E -14\right)$ and $R_s$ is the radius scale at which the first or the second power $\lambda$ dominates. For radius larger than a given radius, $R_{s}$, the density slope is similar, so only one slope $\lambda_{b}$ is considered, allowing for the change in position $R_{s}+R_{s1}\cdot E_{s}$.
 The equation was fitted to proton showers in figure \ref{fig: LDF b} and the parameters obtained for proton and iron showers are displayed in table \ref{tab: LDFfit}.

\begin{table}[h]
\centering{}
\caption[]{Parametrization of the lateral density of Cherenkov photons using equation \ref{eq: LDFfit}, for proton and iron showers.
}
\small
\begin{tabular}{l|lllllll}
\toprule \toprule
primary & $c_0$ & $c_1$ & $\lambda_a$ & $\lambda_{a1}$ & $R_s$ & $R_{s1}$ & $\lambda_b$  \\  
 \tabularnewline  \bottomrule
proton   & $4.86$ & $1.12$ & $-0.396$ & $-0.00985$ & $840$ & $-66.8$ & $-5.21$ \\
iron   & $4.39$ & $1.29$ & $-0.287$ & $-0.0774$ & $940$ & $-49.1$ & $-5.20$ \\
 \tabularnewline
\bottomrule \bottomrule 
\end{tabular}
\label{tab: LDFfit}
\end{table}
\normalsize

\begin{figure}[h]
\begin{center}\centering
                \includegraphics[width=0.65\textwidth]{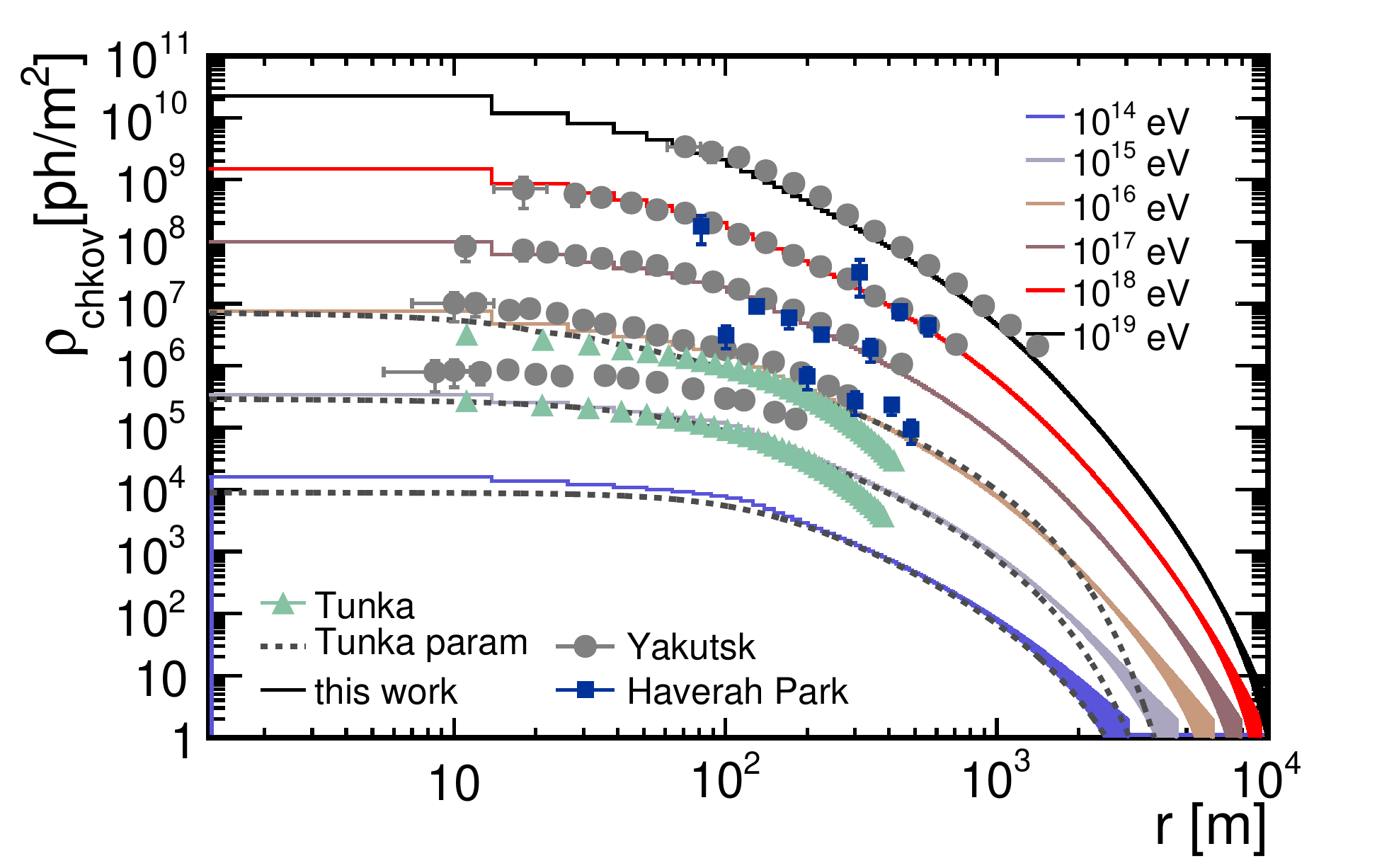}
\caption[]{Photon density distributions, only including photons directions with less than $40^\circ$ with respect to the vertical. The Tunka parametrizations\cite{TunkaExp}, Yakutsk \cite{YakutskExp} and Haverah Park\cite{HaverahPark} results are also drawn. 
}
\label{fig: LDF_OtherExp}
\end{center}
\end{figure}

The results obtained can be compared to other experiments such as Tunka\cite{TunkaExp}, Yakutsk\cite{YakutskExp} and Haverah Park\cite{HaverahPark}. In figure \ref{fig: LDF_OtherExp}, the result obtained for the lateral photon density is compared to available parameterizations and data. These densities corresponds to photons with less than $\sim40^\circ$, with respect to the vertical position, which is approximately the field of view of those experiments. We can see that at higher energies our results are similar to  Yakutsk and Haverah Park. At lower energies, the simulation is similar to Tunka, the differences at $10^{15}$ and $10^{16}$ come mainly from the experiment's energy calibration systematics and detector efficiencies.


\begin{figure}[h]
\begin{center}\centering
       \begin{subfigure}[b]{0.49\textwidth}\centering
                \includegraphics[width=1\textwidth]{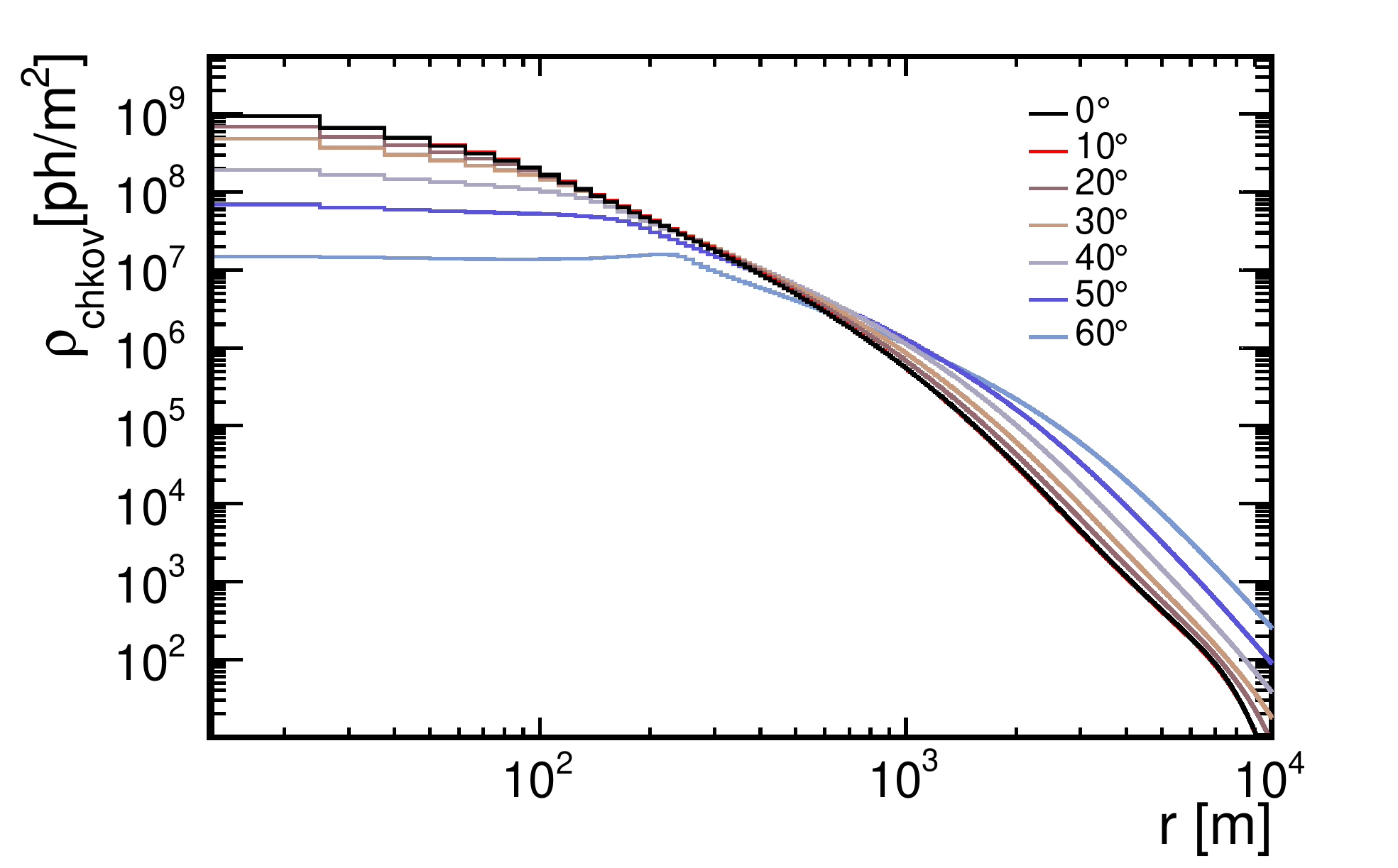}
                \caption{$\rho_{chkov}$ at $10^{18}$ eV}
                \label{fig: LDFtheta a}
        \end{subfigure}%
                \hspace*{0.00\textwidth}
       \begin{subfigure}[b]{0.49\textwidth}\centering
                \includegraphics[width=1\textwidth]{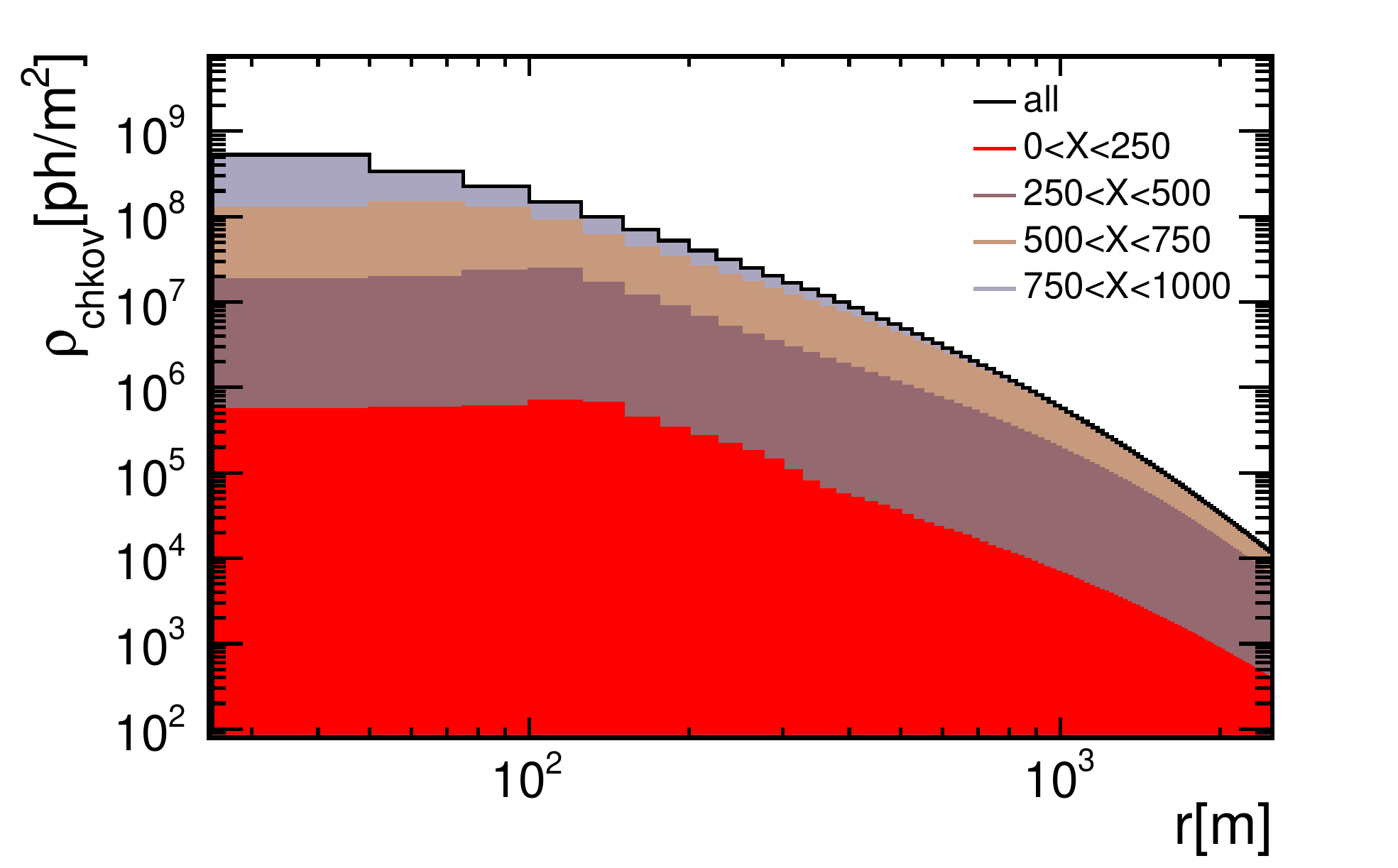}
                \caption{$\rho_{chkov}$ at $10^{18}$ eV and $0^\circ$}
                \label{fig: LDFtheta b}
        \end{subfigure}
\caption[]{a) lateral density of photons for several shower inclinations at $10^{18}$ eV. b) lateral density of photons for several atmospheric depth, where the photons were produced, values for proton showers with $10^{18}$ eV and $0^\circ$.     
}
\label{fig: LDFtheta}
\end{center}
\end{figure}

\begin{figure}[h]
\begin{center}\centering
       \begin{subfigure}[b]{0.49\textwidth}\centering
                \includegraphics[width=1\textwidth]{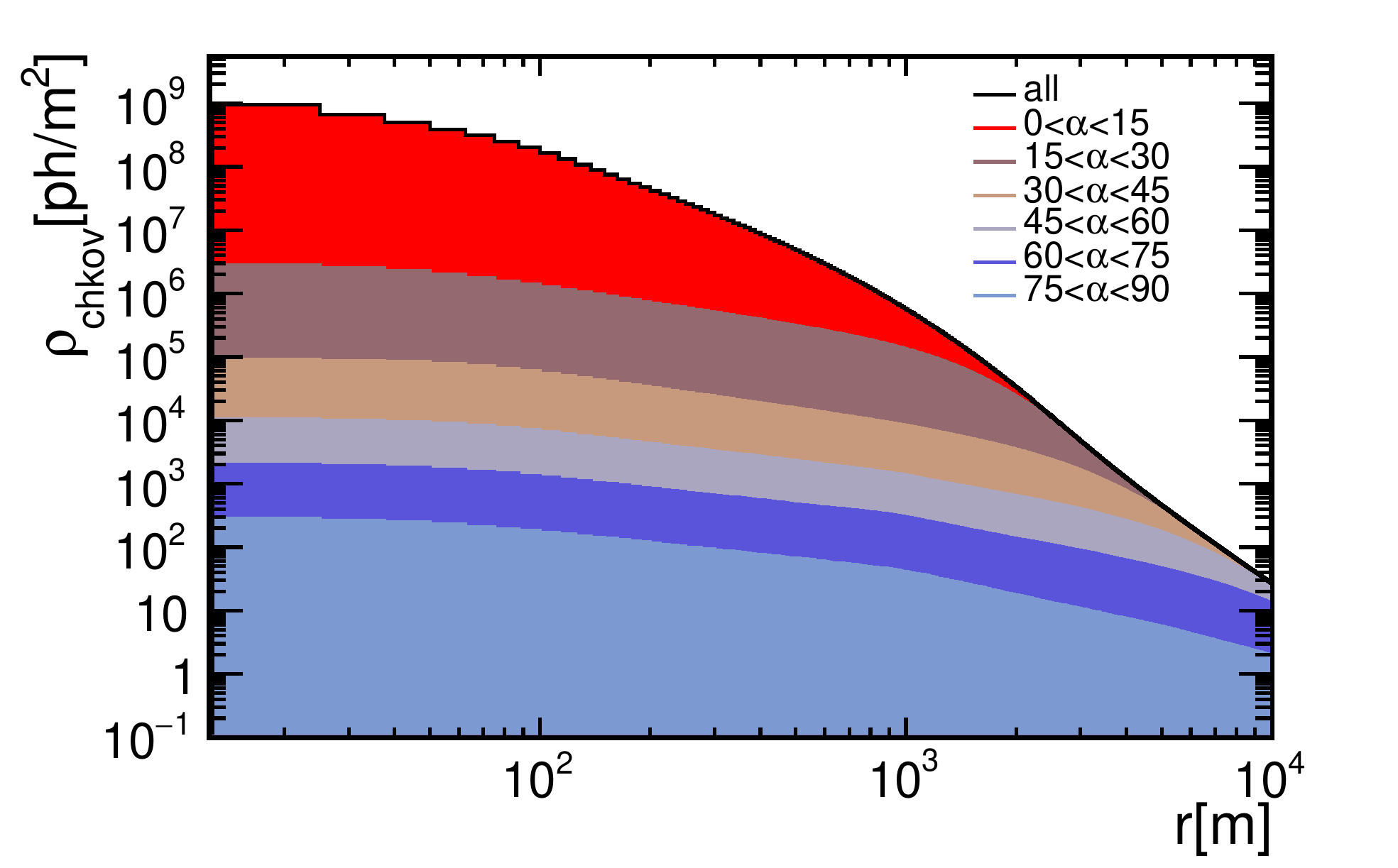}
                \caption{$\rho_{chkov}$ at $0^\circ$}
                \label{fig: LDFalpha a}
        \end{subfigure}%
                \hspace*{0.00\textwidth}
       \begin{subfigure}[b]{0.49\textwidth}\centering
                \includegraphics[width=1\textwidth]{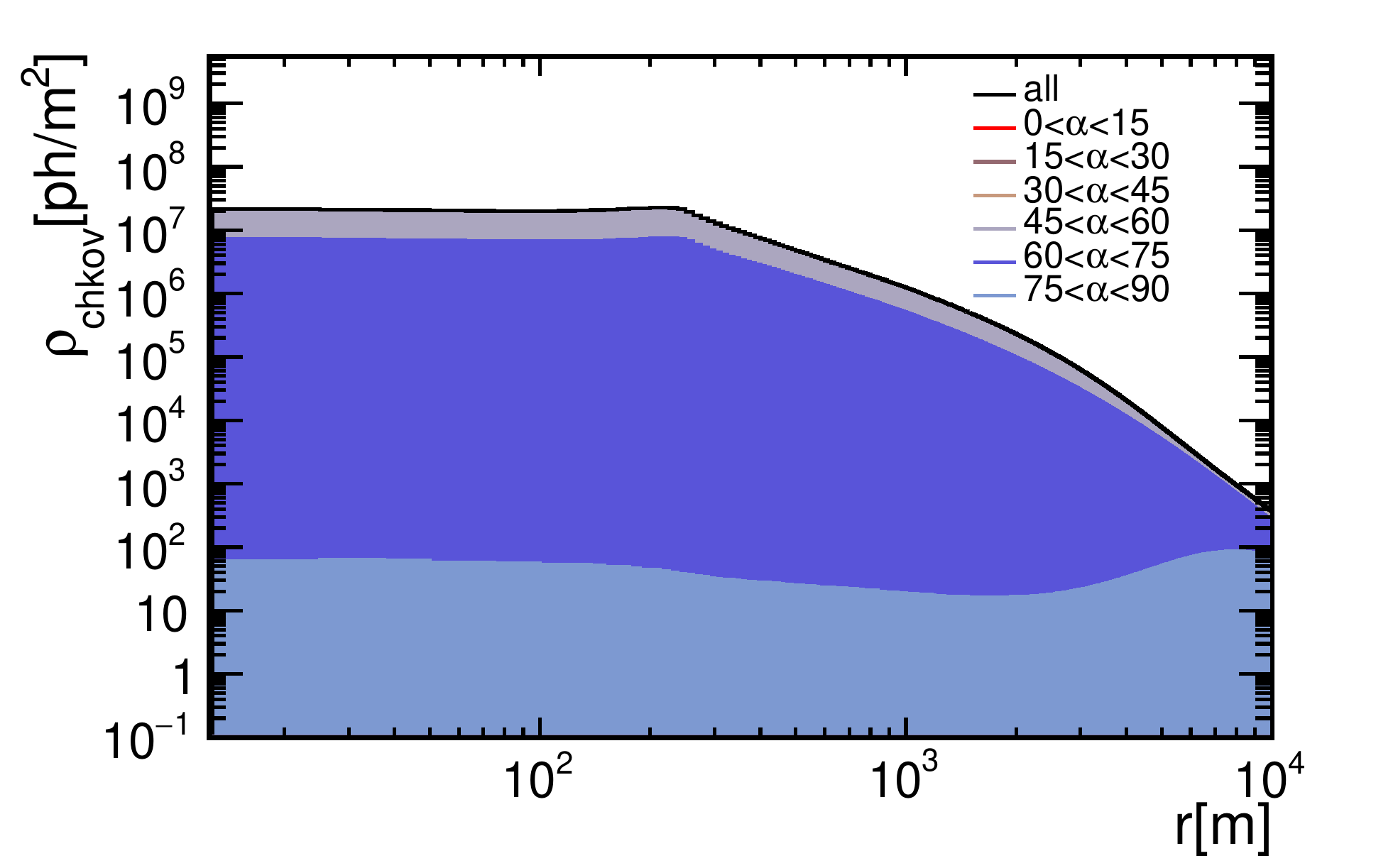}
                \caption{$\rho_{chkov}$ at $60^\circ$}
                \label{fig: LDFalpha b}
        \end{subfigure}
\caption[]{Lateral density of Cherenkov photons as function of the radius for several photon directions with respect to the vertical axis. Distributions for showers with $0^\circ$ (a) and $60^\circ$ (b) at $10^{18}$ eV.
}
\label{fig: LDFalpha}
\end{center}
\end{figure}

In figure \ref{fig: LDFtheta a}, the lateral density of Cherenkov photons is shown for several shower inclinations, at $10^{18}$ eV. The lateral profiles become wider with increasing shower inclination, since the shower will be in a later development stage. Also, in showers with $0^\circ$, at $10^{18}$ eV the longitudinal maximum is close to the ground so the flat shape for radius smaller than $\sim 100$ m is not seen. As the shower gets more inclined, the position of the particle maximum is farther from the ground and the Cherenkov opening angle produces the flat shape. \\
Our simulation also allows us to observe the position in depth, where the photons are produced as functions of the radius, in figure \ref{fig: LDFtheta b}.\\
Cherenkov photons arriving on the ground have different directions with respect to the vertical direction. In previous figures all photons are shown, but it is possible to observe the different directional components. In figures \ref{fig: LDFalpha}, 
the lateral distributions are shown with each photon directional component, for showers with $0^\circ$ (a) and $60^\circ$ (b) at $10^{18}$ eV.

\FloatBarrier
\subsection{Time distributions}
\label{subsection:LDFs}

\begin{figure}[h]
\begin{center}\centering
       \begin{subfigure}[b]{0.49\textwidth}\centering
                \includegraphics[width=1\textwidth]{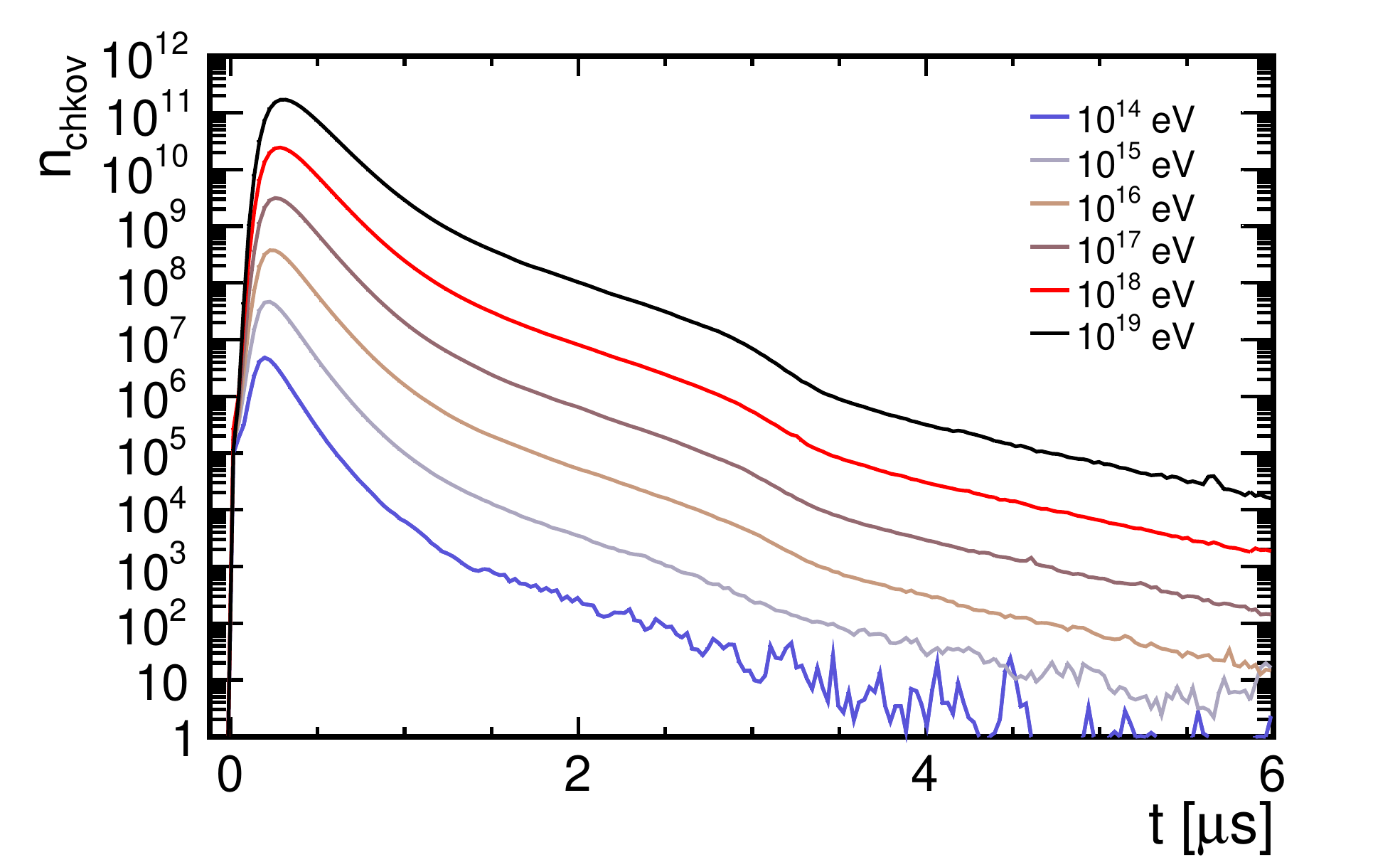}
                \caption{$n_{chkov}$ for $r\in\left[975,1025\right]$ m}
                \label{fig: times a}
        \end{subfigure}%
                \hspace*{0.00\textwidth}
       \begin{subfigure}[b]{0.49\textwidth}\centering
                \includegraphics[width=1\textwidth]{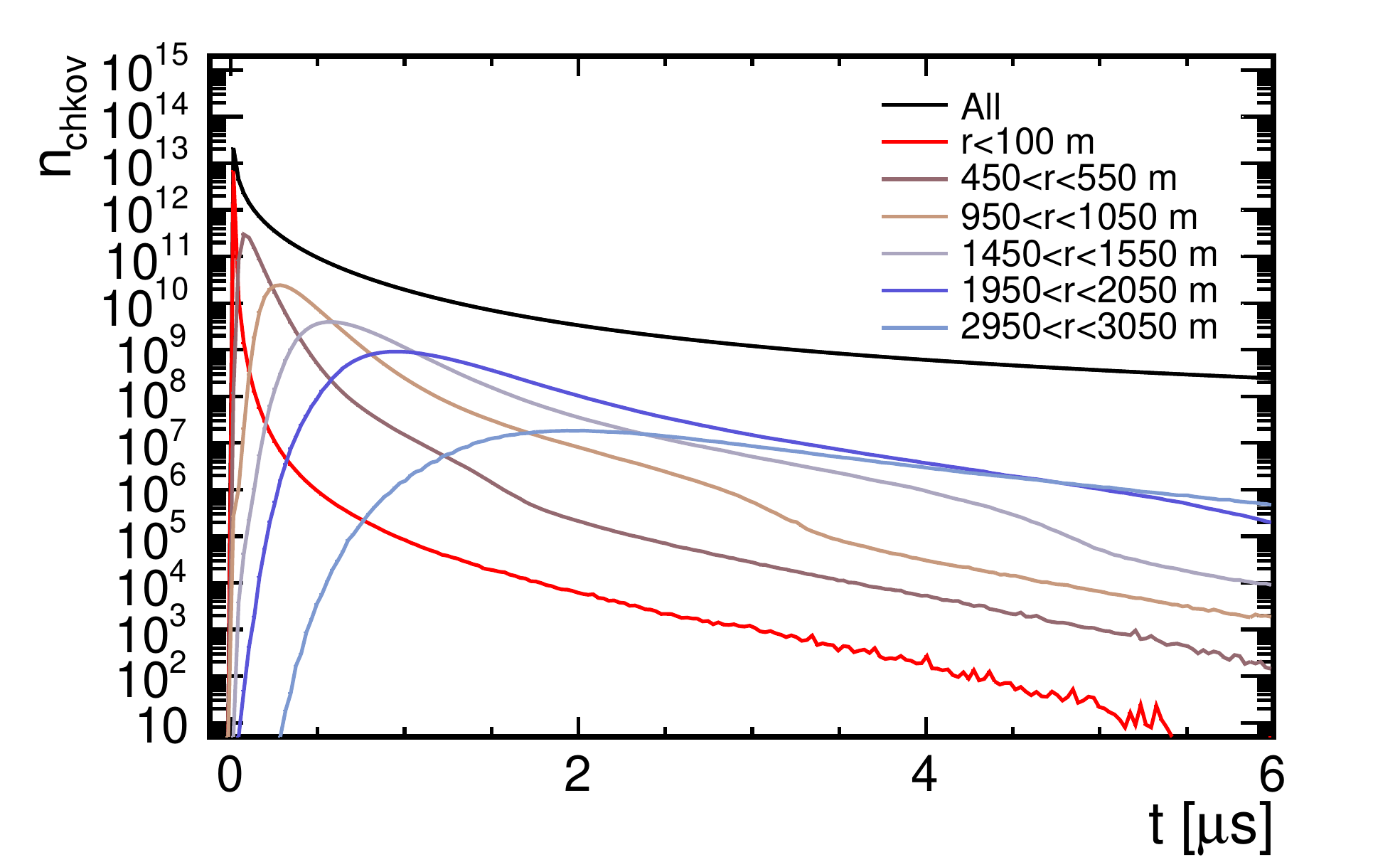}
                \caption{$n_{chkov}$ at $10^{18}$ eV for several r. }
                \label{fig: times b}
        \end{subfigure}
\caption[]{a) number of Cherenkov photons at the ground in $r\in\left[975,1025\right]$ m, for several energies. b) number of photons at $10^{18}$ eV for several distances to the shower axis. Distributions were obtained for vertical proton showers. 
}
\label{fig: times}
\end{center}
\end{figure}

The arrival time of Cherenkov photons at ground is also calculated by our model, as seen in section \ref{subsection:Time}. The final time distribution on the ground, in the range $r\in\left[975,1025\right]$ m, are shown in the figure \ref{fig: times a}, for the same vertical events with several energies. The timing distributions change with the shower energy. 

\begin{figure}[h]
\begin{center}\centering
       \begin{subfigure}[b]{0.49\textwidth}\centering
                \includegraphics[width=1\textwidth]{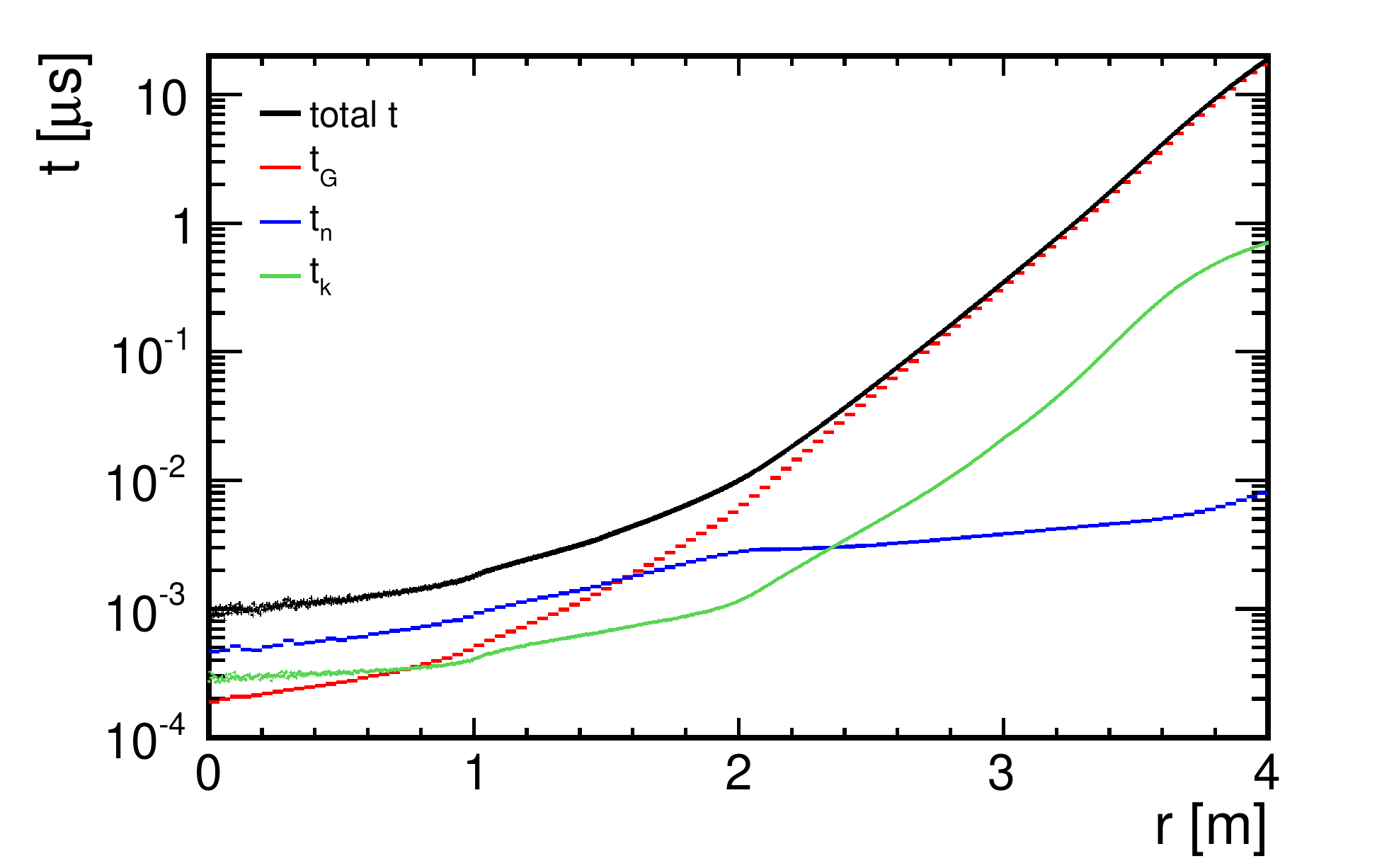}
                \caption{Average time at $10^{18}$ eV and $0^\circ$}
                \label{fig: Avtimes a}
        \end{subfigure}%
                \hspace*{0.00\textwidth}
       \begin{subfigure}[b]{0.49\textwidth}\centering
                \includegraphics[width=1\textwidth]{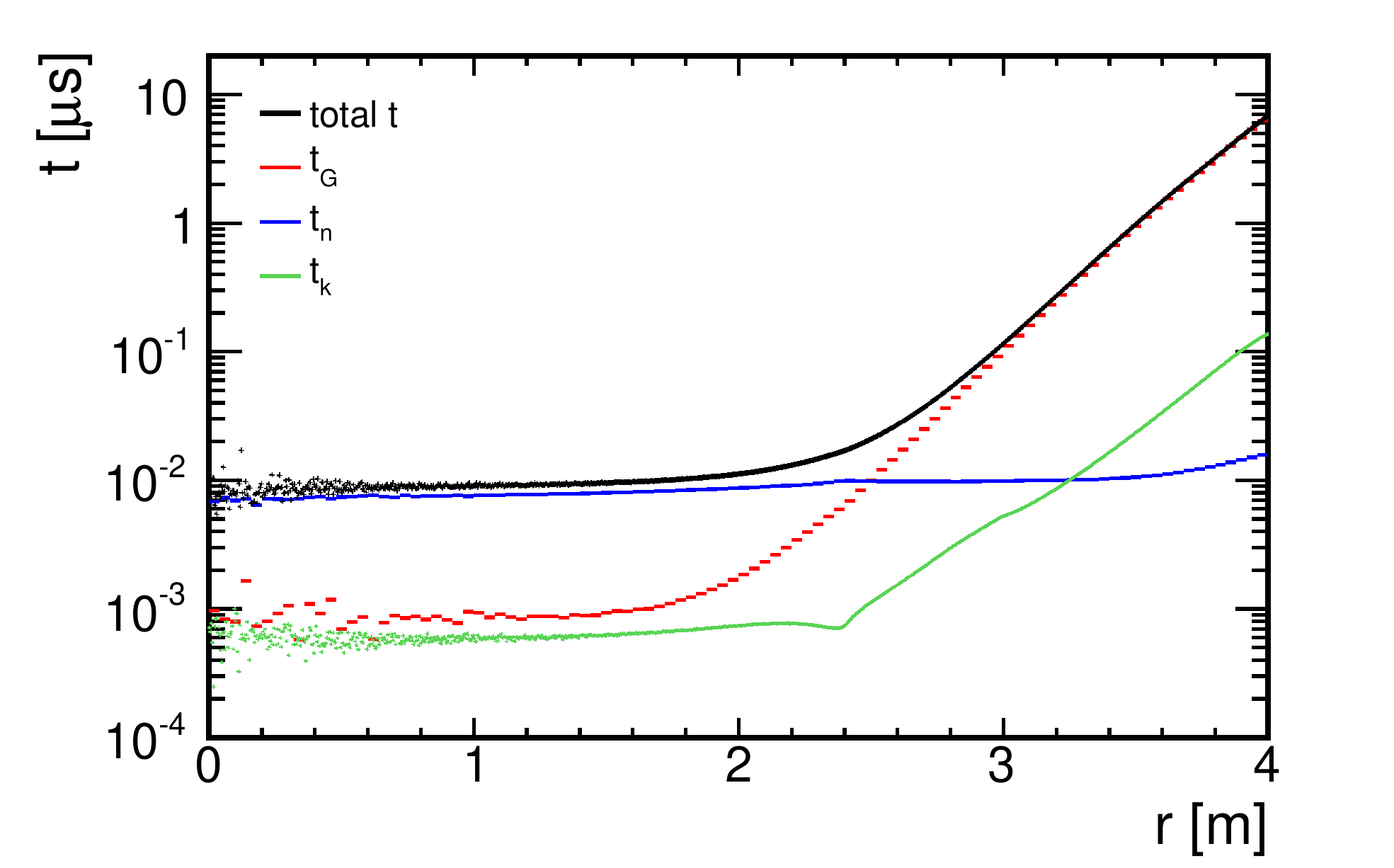}
                \caption{Average time at $10^{18}$ eV and $60^\circ$}
                \label{fig: Avtimes b}
        \end{subfigure}
\caption[]{Different average arrival times of Cherenkov as function of radius to the shower core: total time (black); geometric time delay $t_G$ (red); refractive index time delay $t_n$ (blue); and kinematic time delay $t_k$(green). Values for proton showers with $10^{18}$ eV at $0^\circ$ (a) and $60^{\circ}$ (b).
}
\label{fig: Avtimes}
\end{center}
\end{figure}

The most important feature can be observed in the time distributions for different distances to the shower core. In figure \ref{fig: times b}, we can see that the time distribution changes with the distance to the shower axis. As the distance $r$ increases, the time distribution becomes wider due to the increase of the geometric time delay. \\
The average arrival times as function of the radius are shown in the figures \ref{fig: Avtimes}, with each time contribution shown separately. The different slopes seen for $r<100$ m are due to the Cherenkov opening angle. The delay due to the refractive index grows slowly with $r$, simply due to the increase in the distance travelled by the light. The kinematic delay increases quickly with the radius, since for larger radius, the emission Cherenkov angle increases. This means that the parent particles are not close to the shower axis direction and are less energetic, which corresponds to a larger kinematic delay (it comes from the distribution in figure \ref{fig: Ch_SkyBin d}).\\
It worth to notice that the geometric time delay dominates above $\sim10^{2}$ m. This time if purely geometric and model independent which in the future would allow new analysis.

\begin{figure}[h]
\begin{center}\centering
       \begin{subfigure}[b]{0.49\textwidth}\centering
                \includegraphics[width=1\textwidth]{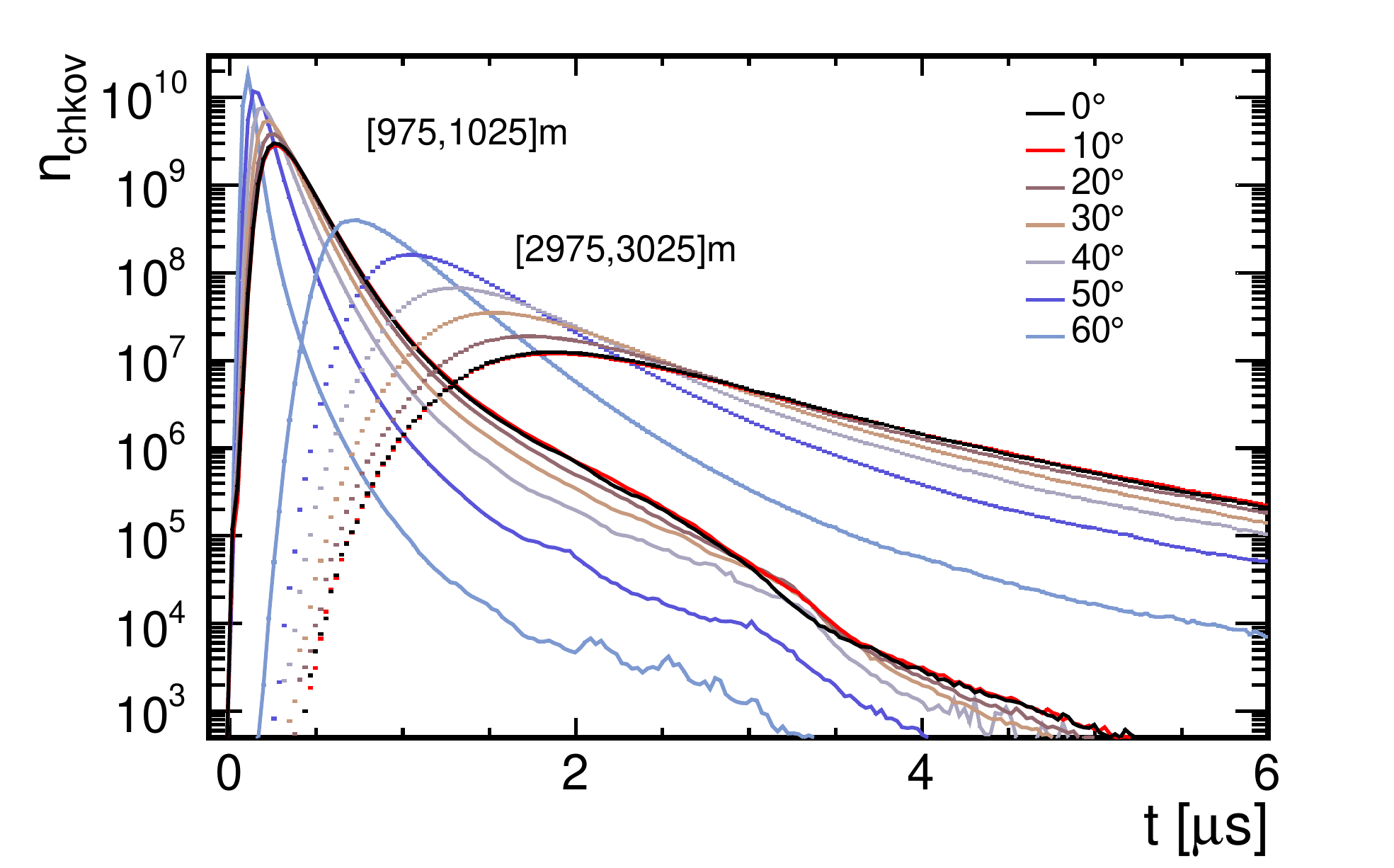}
                \caption{$n_{chkov}$ at $10^{18}$ eV }
                \label{fig: Timetheta a}
        \end{subfigure}%
                \hspace*{0.00\textwidth}
       \begin{subfigure}[b]{0.49\textwidth}\centering
                \includegraphics[width=1\textwidth]{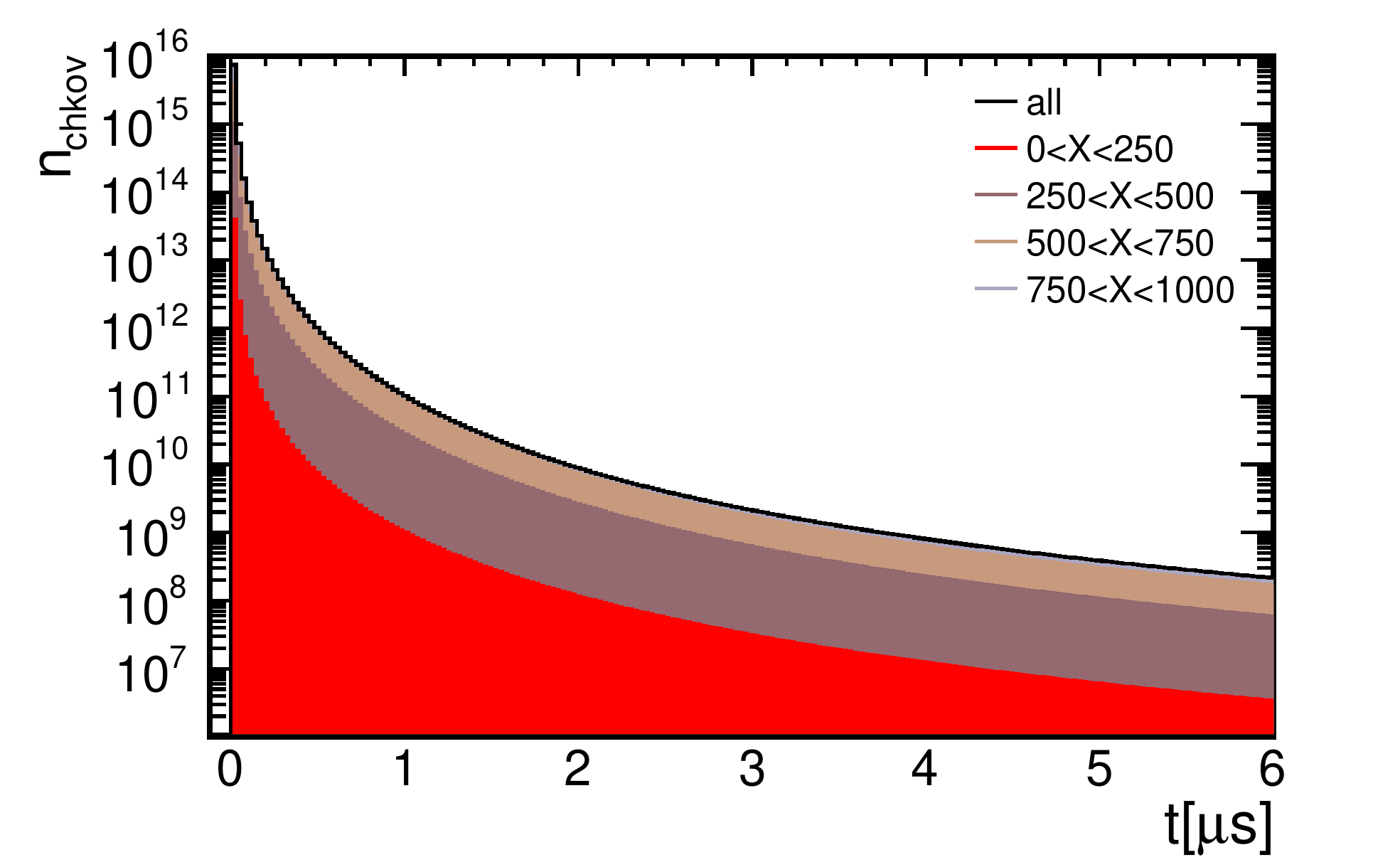}
                \caption{photon time distribution}
                \label{fig: Timetheta b}
        \end{subfigure}
\caption[]{a) time distribution of photons for several shower inclinations, at the radius $r\in\left[975,1025\right]$ and $r\in\left[2925,3025\right]$. b) time distributions of all photons. Each colors corresponds to photons produced at a specific atmospheric depth, values for proton shower with $10^{18}$ eV and $0^\circ$.    
}
\label{fig: Timetheta}
\end{center}
\end{figure}

The effect of the shower inclination in the time distributions can be seen in the figure \ref{fig: Timetheta a}. The distributions are shown at radius of $1000$ m and at $3000$ m, for several shower inclinations with $10^{18}$ eV. It can be seen that, at the same radius in inclined showers, the distributions are narrower and have almost no delay. Particles at a given radius in one inclined shower, correspond to the same particles with lower radius in a vertical shower. So, the geometric time delay is smaller on inclined showers.\\
In figure \ref{fig: Timetheta b}, the time distributions are shown for all photons, with their respective components at the depth where they were produced.

\section{Summary}
\label{section:Conclusions}

This work explores a new approach to simulate Cherenkov light in EAS. With the BinTheSky framework, 3-dimensional information from CORSIKA can be saved and used to simulate Cherenkov light in the atmosphere. The developed method was validated for shower energies around $10^{14}$ and $10^{15}$ eV, for which its results can be compared directly with the CORSIKA output.  At higher energies the  results of the developed method are compatible with Yakutsk measurements. 

Saving Cherenkov photon information in a square of 1500 m around the shower core gives a CORSIKA output per event of around 15 Gb and 170 Gb at $10^{14}$ eV and $10^{15}$ eV respectively. Moreover, we could use the CORSIKA option CERARY to save the Cherenkov in a specific sparse array for distances up to 1500 m and for a few detectors with 1 m$^2$, for example, to save space in the output file. However, this procedure has two disadvantages: on one side we cannot re-simulate the same CORSIKA showers several times, with different core position with respect to the array (which it's a normal procedure in the cosmic ray field, to increase the number of simulations); but the bigger disadvantage is that to calculate the photons emitted along the direction of the array, the CORSIKA program needs to see most of the produced photons, taking more than 75 h to generate one event at $10^{18}$ eV. The 3D simulation takes between $\sim30$ min to $3$ h to simulate all Cherenkov photons within a radius of $10000$ m of the shower core (the time duration depends on the number of filled SkyBins). If a sparse array is considered with the 3D simulation, the requested time easily decreases below 1 min, depending on the array size. 

The 3D simulation allows several new studies on Cherenkov light production, propagation and detection (and also on fluorescence light production, propagation and detection, which can be also considered in the simulation). The framework developed enables the study of the effects of different atmospheric parametrizations and hadronic models on light emission from EAS. For example, in figures \ref{fig: LDFtheta} and \ref{fig: Timetheta} we can see the contributions from different depths for the lateral density and time distributions of Cherenkov photons. The contributions of the different emission directions of Cherenkov photons can be seen in figure \ref{fig: LDFalpha}, whereas the effects of different Mie scattering models on the longitudinal light profile are shown in figure \ref{fig: Validation b}.

\section*{Acknowledgements}
We would like to thank R. Concei\c{c}\~{a}o, Pedro Assis and M. Pimenta for carefully reading this manuscript and comments. The authors wish also to thank FCT-Portugal and CERN/FIS-NUC/0038/2015 for financial support. 

\FloatBarrier
\section*{References}
\bibliography{myref}

\FloatBarrier
\appendix

\end{document}